\documentstyle[11pt]{article}
\begin{document}

\title{ 
{\bf Topics in Finite Temperature Field Theory}}
\author{Ashok Das \\
\\
Department of Physics and Astronomy, \\
University of Rochester,\\
Rochester, New York, 14627.}
\date{}
\maketitle

\begin{abstract}

We discuss a few selected topics in finite temperature field theory.
\end{abstract}

\vfill\eject
\section{Introduction}

Studies of physical systems at finite temperature have led, in the
past,  to many interesting properties such as phase transitions, blackbody
radiation etc. However, the study of complicated quantum mechanical
systems at finite temperature has had a systematic development only in
the past few decades. There are now well developed and well understood
formalisms to describe finite temperature field theories, as they are
called. In fact, as we know now, there are three distinct, but
equivalent formalisms [1-3] to describe such theories and each has its
advantages and disadvantages. But, the important point to note is that
we now have a systematic method of calculating thermal averages
perturbatively in any quantum field theory.

This, of course, has led to a renewed interest in the study of finite
temperature field theories for a variety of reasons. We can now study
questions such as phase transitions involving symmetry restoration in
theories with spontaneously broken symmetry [4]. We can study the
evolution of the universe at early times which clearly is a system at
high temperature. More recently, even questions such as the chiral
symmetry breaking phase transition or the confinement-deconfinement
phase transition in QCD [5-6] have drawn a lot of attention in view of the
planned experiments involving heavy ion collisions. This would help us
understand properties of the quark-gluon plasma better.

The goal of this article is to share, with the readers, some of
the developments in finite temperature field theories in the recent
past and the plan of the article is as follows. In the next section,
we will describe some basic ideas behind describing a quantum mechanical
theory in terms of path integrals [7]. This is the approach which
generalizes readily to the study of finite temperature field
theory. In section {\bf 3}, we will discuss one of the formalisms, in
fact, the oldest one, of describing finite temperature field
theory. This goes under the name of the imaginary time formalism or
the Matsubara formalism [1, 5, 8-10]. In this description, the
dynamical  time is
traded in for the temperature. In contrast, the real time formalisms
of finite temperature field theory contain both time and
temperature. In section {\bf 4}, we discuss one of the real time
formalisms known as thermo field dynamics [3, 10-11]. This is an ideal
description 
to understand operator related issues involving finite temperature
field theories although it has a path integral representation which is
quite nice for calculations as well. The other real time formalism,
which is much older and is known as the closed time path formalism
[2, 10, 12],  is
described in section {\bf 5}. This formalism is very nice because it
describes both equilibrium and non-equilibrium phenomena, at finite
temperature, with equal ease. Temperature leads to many subtle
features in field theories. In section {\bf 6}, we discuss one such
subtlety, namely, how one needs a generalization of the Feynman
combination formula to perform calculations at finite temperature [13]. In
section {\bf 7}, the issue of large gauge invariance is discussed
within the context of a simple quantum mechanical model [14-15]. In section
{\bf 8}, we discuss in some detail how temperature can lead to breaking
of some symmetries like supersymmetry [16] (Temperature normally has the
effect of restoring symmetries). Finally, we present a brief
conclusion in section {\bf 9}. The subject of finite temperature field
theories is quite technical and to
keep the contents simple, we have chosen, wherever possible, simple,
quantum mechanical models to bring out the relevant ideas. Finally, we
would like to note that there are many works in the literature and the
references, at the end, are only representative and are not meant to
be exhaustive in any way.

\section{Path Integrals at Zero Temperature}

In studying a quantum mechanical system or a system described by a
quantum field theory, we are basically interested in determining the
time evolution operator. In the standard framework of quantum
mechanics, one solves the Schr\"{o}dinger equation to determine the
energy eigenvalues and eigenstates simply because the time evolution
operator is related to the Hamiltonian. There is an alternate method
for evaluating the matrix elements of the time evolution operator
which is useful in studying extremely complicated physical
systems. This goes under the name of path integral formalism [7, 17-18].

In stead of trying to develop the ideas of the path integral
formalism here, let us simply note that, for a bosonic system described by
a time independent quantum mechanical Hamiltonian, the transition
amplitude can be represented as (The subscript $H$ denotes the
Heisenberg picture.)
\begin{equation}
_{H}\langle x_{f},t_{f}|x_{i},t_{i}\rangle_{H} = \langle x_{f}|e^{-{i\over
\hbar}H(t_{f}-t_{i})}|x_{i}\rangle =  \int {\cal D}x\,e^{{i\over
\hbar}S[x]}\label{b1}
\end{equation}
There are several comments in order. First, the transition amplitude
is nothing other than the matrix element of the time evolution
operator in the coordinate basis. Second, the integral on the right
hand side is known as a path integral. It is an integral over all
possible paths connecting the initial coordinate $x_{i}$ and the final
coordinate $x_{f}$ which are held fixed. The simplest way to evaluate
such an integral is to divide the time interval of the path
between $x_{i}$ and $x_{f}$ into $N$ intervals of equal length. Integrating
over all possible values of the coordinates of the  intermediate
points  (which are ordinary integrals) and taking $N\rightarrow
\infty$  such that the time interval is held fixed is equivalent to
integrating over all possible paths. Finally, the action $S[x]$ in the
exponent of the integrand is nothing other than the classical action
for the bosonic system under study. This is true for most conventional
physical systems where the Hamiltonian depends quadratically on the
momentum. If this is not the case (and there are some cases where it
is not), the right hand side of (\ref{b1}) needs to be
modified. However, for most systems that we will discuss, we do not
have to worry about this fine point.

The advantage of the path integral is that while the left hand side
involves quantum mechanical operators, the right hand side is
described only in terms of classical variables and, therefore, the
manipulations become quite trivial. Furthermore, the transition
amplitude defined in eq. (\ref{b1}) can be generalized easily to
incorporate sources  and this allows us to derive  various Greens
functions of the
theory in a very simple and straightforward manner. As an example, let
us simply note here that for a harmonic oscillator, the action is
quadratic in the dynamical variables, namely,
\[
S[x] = \int_{t_{i}}^{t_{f}} dt\,\left[{1\over 2}m\dot{x}^{2} - {1\over
2}m\omega^{2}x^{2}\right]
\]
and, in this case, the path integral can be exactly evaluated and has
the form [7]
\begin{eqnarray}
\langle x_{f}|e^{-{i\over \hbar}HT}|x_{i}\rangle & = & \int {\cal
D}x\,e^{{i\over \hbar}S[x]}\nonumber\\
 & = & \left({m\omega\over 2\pi i\hbar\sin\omega T}\right)^{{1\over
2}}\,e^{{i\over \hbar}S[x_{cl}]}\label{b2}
\end{eqnarray}
Here, we have defined $T=t_{f}-t_{i}$. $S[x_{cl}]$ represents the
action associated with the classical trajectory (satisfying the
Euler-Lagrange equation) and has the form
\begin{equation}
S[x_{cl}] = {m\omega\over 2\sin\omega
T}\left[(x_{i}^{2}+x_{f}^{2})\cos\omega T -
2x_{i}x_{f}\right]\label{b3}
\end{equation}

The path integrals can also be extended to quantum mechanical systems
describing fermionic particles. However, one immediately recognizes
that there are no classical variables which are fermionic. Therefore,
in order to have a path integral description of such systems in terms
of  classical
variables, we must supplement our usual notions of classical variables
with anti-commuting Grassmann variables [19]. With this, for example, we
can write a classical action for the fermionic oscillator as
\begin{equation}
S[\psi,\bar{\psi}] = \int_{t_{i}}^{t_{f}} dt\,(i\bar{\psi}\dot{\psi} -
\omega\bar{\psi}\psi)\label{b4}
\end{equation}
Here $\psi$ and $\bar{\psi}$ are anti-commuting Grassmann variables
and in the quantum theory, as operators, can be identified with the
fermionic annihilation and creation operators respectively. The action
in eq. (\ref{b4}) is also quadratic in the variables much like the
bosonic oscillator and the path integral for the fermionic oscillator
can also be exactly evaluated giving [7]
\begin{eqnarray}
\langle \psi_{f},\bar{\psi}_{f}|e^{-{i\over
\hbar}HT}|\psi_{i},\bar{\psi}_{i}\rangle & = & \int {\cal
D}\bar{\psi}\,{\cal D}\psi\,e^{{i\over
\hbar}S[\psi,\bar{\psi}]}\nonumber\\
 & = & e^{{i\omega T\over 2}}\,e^{\left(e^{-i\omega
T}\bar{\psi}_{f}\psi_{i}-\bar{\psi}_{f}\psi_{f}\right)}\label{b5}
\end{eqnarray}

In a quantum field theory, we are often interested in evaluating time
ordered correlation functions in the vacuum because the S-matrix elements
can be obtained from such Greens functions. These can be derived in a
natural manner from what is known as the vacuum to vacuum transition
functional which can be obtained from the transition amplitude in
eq. (\ref{b1}) in a simple manner and also has a path integral
representation of the form  
\begin{equation}
\lim_{T\rightarrow\infty}\langle 0|e^{-{i\over \hbar}HT}|0\rangle =
\int {\cal D}x\,e^{{i\over \hbar}S[x]}\label{b6}
\end{equation}
where
\begin{equation}
S[x] = \int_{-\infty}^{\infty} dt\,L(x,\dot{x})\label{b7}
\end{equation}
Furthermore, the path integral in eq. (\ref{b6}) has no end-point
restriction unlike in eq. (\ref{b1}). This vacuum to vacuum transition
amplitude is also commonly denoted by $\langle 0|0\rangle$ with the
limiting process understood. We note here that an
analogous formula also holds for fermionic systems.

The vacuum to vacuum amplitude in the presence of a source has the
form
\begin{equation}
Z[J] = \langle 0|0\rangle_{J} = \int {\cal D}x\,e^{{i\over
\hbar}S[x,J]}\label{b8}
\end{equation}
where
\begin{equation}
S[x,J] = S[x] + \int_{-\infty}^{\infty} dt\,J(t)x(t)\label{b9}
\end{equation}
Here $J(t)$ is a classical source and it can be easily checked that,
in the limit of vanishing source, the functional derivatives of $Z[J]$
give rise  to time ordered Greens functions in the vacuum.

With this very brief review of the path integral description for zero
temperature quantum mechanical theories, we are now ready to describe
the different formalisms available to study quantum mechanical systems
at finite temperature.

\section{Imaginary Time Formalism}

The properties of a  quantum mechanical system, at finite temperature,
can  also be given a path integral description. There are various, but
equivalent ways of doing this. Of the different formalisms available
to study a  quantum mechanical system at finite temperature, the
imaginary time  formalism is the oldest [1]. To appreciate this, let us
recall some of the features of a   statistical ensemble. A statistical
ensemble in equilibrium at a finite temperature ${1\over \beta}$ (in
units of Boltzmann constant) is described in terms of a partition
function 
\begin{equation}
Z(\beta) = {\rm Tr}\,\rho(\beta) = {\rm Tr}\,e^{-\beta {\cal H}}\label{c1}
\end{equation}
Here $\rho(\beta)$ is known as the density matrix (operator) and
${\cal H}$ can be thought of as the generalized Hamiltonian of the
system. If
\[
{\cal H} = H
\]
where $H$ is the Hamiltonian of the system, we say that the
ensemble is a canonical ensemble where the particle number is fixed
and the system is allowed to exchange only energy with a heat bath. On
the other hand, if
\[
{\cal H} = H - \mu\,N
\]
where $N$ is the number operator, then, the ensemble is known as a
grand canonical ensemble where the system can exchange not only energy
with a heat bath, but can also exchange particles with a
reservoir. The  constant
$\mu$ is known as the chemical potential. In a statistical ensemble,
of course, the important observables are the ensemble averages and,
for any observable ${\cal O}$, they are defined as
\begin{equation}
\langle {\cal O}\rangle_{\beta} = {1\over Z(\beta)}\,{\rm
Tr}\,\rho(\beta){\cal O}\label{c2}
\end{equation}
Let us also note here that since the partition function involves a
trace, it leads to an interesting identity following from the
cyclicity of the trace, namely, (we will assume from now on, unless
otherwise specified, that $\hbar = 1$)
\begin{eqnarray}
\langle {\cal O}_{1}(t){\cal O}_{2}(t')\rangle_{\beta} & = & {1\over
Z(\beta)}\,{\rm Tr}\,e^{-\beta{\cal H}}{\cal O}_{1}(t){\cal
O}_{2}(t')\nonumber\\
 & = & {1\over Z(\beta)}\,{\rm Tr}\,e^{-\beta{\cal H}}{\cal
O}_{2}(t')e^{-\beta{\cal H}}{\cal O}_{1}(t)e^{\beta{\cal
H}}\nonumber\\
 & = & {1\over Z(\beta)}\,{\rm Tr}\,e^{-\beta{\cal H}}{\cal
O}_{2}(t'){\cal O}_{1}(t+i\beta)\nonumber\\
 & = & \langle {\cal O}_{2}(t'){\cal O}_{1}(t+i\beta)\rangle_{\beta}\label{c3}
\end{eqnarray}
Such a relation is known as the KMS (Kubo-Martin-Schwinger) [20] relation
which generalizes to all statistical ensemble averages and plays a
crucial role in the study of finite temperature field theories.

It was observed quite early by Bloch [21] that the operator $e^{-\beta{\cal
H}}$ in the definition of the partition function is like the time evolution
operator in the imaginary time axis. This  is really at the heart of the
imaginary time formalism. In fact, let us note that the canonical
partition function can be written as (with the trace taken in the
coordinate basis)
\begin{equation}
Z(\beta) = \int dx\,\langle x|e^{-\beta H}|x\rangle\label{c4}
\end{equation}
It is clear now that if we identify $T=-i\beta$ in eq. (\ref{b1}),
then, we can give the partition function a path integral
representation as ($\hbar=1$)
\begin{equation}
Z(\beta) = \int {\cal D}x\,e^{-S_{E}[x]}\label{c5}
\end{equation}
where $S_{E}[x]$ is the Euclidean (imaginary time) action for the
system defined over a finite time interval as
\begin{equation}
S_{E}[x] = \int_{0}^{\beta} dt\,L_{E}(x,\dot{x})\label{c6}
\end{equation}
Furthermore, it is clear from eq. (\ref{c4}) that the variable $x$
must satisfy the periodic boundary condition
\begin{equation}
x(\beta) = x(0)\label{c7}
\end{equation}
for eq. (\ref{c5}) to represent a trace (namely, the initial and the
final states must be the same) and that the end point is being
integrated over in the path integral in eq. (\ref{c5}) unlike in
eq. (\ref{b1}). (It is important to note that the original work of
Matsubara  is an operator description of the imaginary time, but we
will  not discuss it in the present article.)

In fact, as an example, let us evaluate the canonical partition
function for the bosonic oscillator using this formalism [7]. The
transition amplitude is already given for zero temperature in
eq. (\ref{b2}). Now making the identifications
\begin{equation}
T=-i\beta,\qquad x_{i}=x_{f}=x\label{c8}
\end{equation}
we obtain from eqs. (\ref{b2}) and (\ref{c4})
\begin{eqnarray}
Z(\beta) & = & \int dx\,\left({m\omega\over
2\pi\sinh\beta\omega}\right)^{{1\over
2}}\,e^{-(m\omega\tanh{\beta\omega\over 2})\,x^{2}}\nonumber\\
 & = & \left({m\omega\over 2\pi\sinh\beta\omega}\right)^{{1\over
2}}\left({\pi\over m\omega\tanh{\beta\omega\over 2}}\right)^{{1\over
2}}\nonumber\\
 & = & {e^{{\beta\omega\over 2}}\over e^{\beta\omega}-1}\label{c9}
\end{eqnarray}
This is, indeed, the partition function for the bosonic oscillator as
can be directly verified.

The partition function, for a fermionic system, can also be similarly
given a path integral representation. However, the anti-commuting
nature of the fermion variables introduces one crucial difference,
namely, for a fermion theory, we have
\begin{equation}
Z(\beta) = \int {\cal D}\bar{\psi}{\cal
D}\psi\,e^{-S_{E}[\psi,\bar{\psi}]}\label{c10}
\end{equation}
with anti-periodic boundary conditions [10]
\begin{equation}
\psi(\beta) = -\psi(0),\qquad \bar{\psi}(\beta) =
-\bar{\psi}(0)\label{c11}
\end{equation}
The Euclidean (imaginary time) action is again defined over a finite
time interval as in eq. (\ref{c6}). In fact, let us calculate the
canonical partition function for a fermionic oscillator, as an
example, from the  result
in eq. (\ref{b5}) as well as the identifications in (\ref{c11}) [7]. Using
\[
\psi_{f} = - \psi_{i} = -\psi,\qquad \bar{\psi}_{f} = -\bar{\psi}_{i}
= -\bar{\psi}
\]
we obtain (remember $T=-i\beta$)
\begin{eqnarray}
Z(\beta) & = & \int d\bar{\psi}d\psi\,e^{{\beta\omega\over
2}}\,e^{-(1+e^{-\beta\omega})\bar{\psi}\psi}\nonumber\\
  & = & e^{{\beta\omega\over 2}}\,(1+e^{-\beta\omega}) =
2\cosh{\beta\omega\over 2}\label{c12}
\end{eqnarray}
In evaluating this, we have made use of the Berezin rules of
integration [19] for Grassmann variables and we note that
eq. (\ref{c12}),  indeed, gives the correct partition  function for a
fermionic oscillator as can be directly calculated.

Although our discussion so far has been within the context of
simple quantum mechanical systems, everything we have said can be carried over
to a quantum field theory. The partition function for a quantum field
theory can again be written as a path integral involving a Euclidean
action as
\begin{equation}
Z(\beta) = \int {\cal D}\bar{\psi}{\cal D}\psi{\cal
D}\phi\,e^{-S_{E}[\phi,\psi,\bar{\psi}]}\label{c13}
\end{equation}
where the Euclidean action is defined over a finite time interval and
the fields satisfy the periodicity (anti-periodicity)
conditions\begin{equation}
\phi(\beta,\vec{x}) = \phi(0,\vec{x}),\qquad \psi(\beta,\vec{x}) =
-\psi(0,\vec{x})\label{c14}
\end{equation}
and so on. The discussion is slightly more involved for gauge theories and to
keep things simple, we will not discuss gauge theories.

This formulation of a field theory at finite temperature is known as
the imaginary time formalism or the Matsubara formalism [1] and is the
oldest formalism.  There are several distinguishing features of this
formalism. For example, since the time interval is finite, Fourier
transformation  of the time variable would involve discrete
energies. In other words,  the Fourier transform of the propagator,
say for  example, at finite temperature in the imaginary time
formalism, would  take the general form
\begin{equation}
{\cal G}_{\beta}(\tau,\vec{x}) = {1\over
\beta}\sum_{n}e^{-i\omega_{n}\tau}{\cal
G}_{\beta}(\omega_{n},\vec{x})\label{c15}
\end{equation}
where $\omega_{n}={n\pi\over \beta}$ with $n=0,\pm 1,\pm
2,\cdots$. However, from the definition of the time ordered
product 
\begin{equation}
T_{\tau}(\phi(\tau)\phi^{\dagger}(\tau')) = \theta(\tau
-\tau')\phi(\tau)\phi^{\dagger}(\tau')\pm\theta(\tau'
-\tau)\phi^{\dagger}(\tau')\phi(\tau)\label{c16}
\end{equation}
where we have allowed for both bosonic and fermionic fields and the
KMS condition  in eq. (\ref{c3}), it follows that, for $\tau<0$,
\begin{equation}
{\cal G}_{\beta}(\tau,\vec{x}) = \pm {\cal
G}_{\beta}(\tau+\beta,\vec{x})\label{c17})
\end{equation}
It is important to recognize that the periodicity (anti-periodicity)
of the propagator arises from the definition of the time ordered
product for the bosonic (fermionic) fields and the KMS condition and
is not directly connected with the periodicity (anti-periodicity) of
the corresponding field variables which we have discussed
earlier. This periodicity (anti-periodicity) of the propagator,
on the other hand, leads to the restriction that eq. (\ref{c15}) holds with
\begin{equation}
\omega_{n} = \left\{\begin{array}{cl}
                    {2n\pi\over \beta} & {\rm for\; bosons}\\
                    {(2n+1)\pi\over \beta} & {\rm for\; fermions}
                    \end{array}\right.\label{c18}
\end{equation}
where $n=0,\pm 1,\cdots$. These are conventionally known as the
Matsubara frequencies [22].

Given this, one can now calculate the propagators for bosonic and
fermionic field theories in the Matsubara formalism and they take the
forms (in the momentum space)
\begin{eqnarray}
{\cal G}_{\beta}(\omega_{n},\vec{k}) & = & {1\over
\omega_{n}^{2}+\vec{k}^{2}+m^{2}} = {1\over
({2n\pi\over\beta})^{2}+\vec{k}^{2}+m^{2}}\label{c19}\\
  &  & \nonumber\\
{\cal S}_{\beta}(\omega_{n},\vec{k}) & = &
{\gamma^{0}\omega_{n}+\vec{\gamma}\cdot\vec{k}+m\over
\omega_{n}^{2}+\vec{k}^{2}+m^{2}} =
{\gamma^{0}({(2n+1)\pi\over\beta})+\vec{\gamma}\cdot\vec{k}+m\over
({(2n+1)\pi\over\beta})^{2}+\vec{k}^{2}+m^{2}}\label{c20}
\end{eqnarray}

Perturbative calculations can now be developed quite analogously to
the zero temperature field theory. For example, given a field theory,
we can read out the vertices from the Euclidean form of the action and
use the propagators of eq. (\ref{c19}, \ref{c20}) to carry out a
diagrammatic calculation which would lead to the ensemble average for a
given observable. It is clear  that, because the time interval is
finite in this formalism, the coordinate space calculation of any
diagram is cumbersome. However, much like at zero temperature, the
momentum space calculation is much simpler. However, one should keep
the difference in mind, namely, that, at finite temperature, the
external and the internal energies are discrete as in
eq. (\ref{c18}). Consequently, the integration over internal energies
(of zero temperature) is replaced by a sum over the internal
energies. More specifically, we must use
\begin{equation}
\int {d^{4}k\over (2\pi)^{4}}\;\longrightarrow {1\over
\beta}\sum_{n}\int {d^{3}k\over (2\pi)^{3}}\label{c21}
\end{equation}

As an example, let us consider the self-interacting scalar theory
described by
\begin{equation}
{\cal L} = {1\over 2}\partial_{\mu}\phi\partial^{\mu}\phi -{m^{2}\over
2}\phi^{2}-{\lambda\over 4!}\phi^{4}\qquad \lambda>0\label{c22}
\end{equation}
We note that the only one loop correction in this theory is the mass
correction. Rotating to Euclidean space and using the propagator for a
scalar theory as given in eq. (\ref{c19}) as well as (\ref{c21}), we
obtain the one loop mass correction to be
\begin{eqnarray}
\Delta m^{2} & = & {\lambda\over 2\beta}\sum_{n}\int {d^{3}k\over
(2\pi)^{3}}\;{1\over ({2n\pi\over
\beta})^{2}+\vec{k}^{2}+m^{2}}\nonumber\\
 & = & {\lambda\over 2\beta}\left({\beta\over
2\pi}\right)^{2}\sum_{n}\int {d^{3}k\over (2\pi)^{3}}\;{1\over
n^{2}+({\beta\omega_{k}\over 2\pi})^{2}}\label{c23}
\end{eqnarray}
Here, we have introduced the notation,
\begin{equation}
\omega_{k} = (\vec{k}^{2} + m^{2})^{{1\over 2}}\label{c24}
\end{equation}
The sum, in eq. (\ref{c23}), can be easily evaluated using the method
of residues leading to
\begin{equation}
\sum_{n=-\infty}^{\infty}\;{1\over n^{2}+y^{2}} = {\pi\over y}\coth
\pi y\qquad {\rm for}\;y>0\label{c25}
\end{equation}
Using this, the one loop mass correction can be determined to be [23]
\begin{eqnarray}
\Delta m^{2} & = & {\lambda\over 4}\int {d^{3}k\over
(2\pi)^{3}}\;{1\over \omega_{k}}\,\coth\left({\beta\omega_{k}\over
2}\right)\nonumber\\
 & = & {\lambda\over 4}\int {d^{3}k\over (2\pi)^{3}}\,{1\over
\omega_{k}} + {\lambda\over 2}\int {d^{3}k\over (2\pi)^{3}}\,{1\over
\omega_{k}}\,{1\over e^{\beta\omega_{k}}-1}\nonumber\\
 & = & \Delta m_{0}^{2} + \Delta m_{\beta}^{2}\label{c26}
\end{eqnarray}

There are several things to note from this calculation. First, the
mass correction separates into two parts -- one independent of
temperature and the other genuinely a finite temperature
correction. The temperature independent part (zero temperature part)
is divergent as is expected at zero temperature and the divergence has
to be handled by the usual process of renormalization. However, the
finite temperature part is completely free from ultraviolet
divergence. This is a general feature of finite temperature field
theories that temperature does not introduce any new ultraviolet
divergence. We will return to this question later within the context
of  real time formalisms for finite temperature field theories. Let
us also note that (see (\ref{c26})) the finite temperature integrals
are, in general hard to evaluate and cannot be evaluated in a closed
form. However, we can always make a high temperature expansion (small
$\beta$) which would give the temperature dependent correction to the
mass as
\begin{equation}
\Delta m_{\beta}^{2} \approx {\lambda\over 24\beta^{2}} = {\lambda
T^{2}\over 24}\label{c27}
\end{equation}  
This shows that temperature induces a mass correction which is
positive. Intuitively, it is clear that this is the behavior we would
expect from a particle moving in a medium and, furthermore, the
positivity of this correction is crucial in the study of symmetry
restoration in field theories with spontaneous symmetry breaking.

This gives a flavor of calculations at finite temperature,
particularly, in the imaginary time (Matsubara) formalism. It is worth
noting here that, by construction, the imaginary time formalism would
describe physical systems in equilibrium quite well. Since we have
traded the time variable for temperature, it is well suited to
calculate static, equilibrium quantities. Slow temperature dependence
can, however, be brought in by analytically rotating the final result
to Minkowski time [24]. This rotation is, on the other hand, nontrivial
since we only have information about quantities at discrete energy
values in the Euclidean space. The imaginary time formalism is not
suitable to discuss non-equilibrium phenomena.

\section{Thermo Field Dynamics}

As we have seen, in the imaginary time formalism, the time variable is
traded for the temperature. However, in studying various processes, it
is desirable to have the time coordinate in addition to the
temperature. Formalisms where this can be achieved are known as the real
time formalisms and there are two distinct, but equivalent such
formalisms. In this section, we will discuss the formalism of
thermo field dynamics [3, 11, 25] returning to the alternate formalism
in the next section. 

Let us recall from (\ref{c2}) that the ensemble average of any
observable is given by
\begin{eqnarray}
\langle {\cal O}\rangle_{\beta} & = & {1\over Z(\beta)}\,{\rm Tr}\,e^{-\beta
{\cal H}}\,{\cal O}\nonumber\\
 & = & {1\over Z(\beta)}\,\sum_{n} e^{-\beta E_{n}}\,\langle n|{\cal
O}|n\rangle\label{d1}
\end{eqnarray}
Here, we have assumed that the eigenvalues of ${\cal H}$ are discrete,
for simplicity, and that
\begin{eqnarray}
{\cal H}|n\rangle & = & E_{n}|n\rangle\nonumber\\
\langle m|n\rangle & = & \delta_{mn}\nonumber\\
\sum_{n} |n\rangle\langle n| & = & I\label{d2}
\end{eqnarray}

At zero temperature, we know that the Feynman diagrams correspond to
vacuum expectation values of time ordered products. Thus, intuitively,
it is clear that if we can express the ensemble averages as
expectation values in some vacuum (say, a thermal vacuum), then, we
can take over  all the diagrammatic machinery of the zero temperature
field theory. The question, therefore, is whether we can define a vacuum,
say $|0,\beta\rangle$, such that we can write any ensemble average as
\begin{equation}
\langle {\cal O} \rangle_{\beta} = \langle 0,\beta|{\cal
O}|0,\beta\rangle = {1\over Z(\beta)}\,\sum_{n} e^{-\beta
E_{n}}\,\langle n|{\cal O}|n\rangle\label{d3}
\end{equation}

Let us suppose that we can define such a thermal vacuum state as a
linear superposition of the states in our physical Hilbert space, namely,
\begin{equation}
|0,\beta\rangle = \sum_{n}|n\rangle\langle n|0,\beta\rangle = \sum_{n}
 f_{n}(\beta)|n\rangle\label{d4}
\end{equation}
This would lead to
\begin{equation}
\langle 0,\beta|{\cal O}|0,\beta\rangle = \sum_{n,m}
f_{n}^{*}(\beta)f_{m}(\beta)\,\langle n|{\cal O}|m\rangle\label{d5}
\end{equation}
Consequently, this would coincide with eq. (\ref{d3}) only if
\begin{equation}
f_{n}^{*}(\beta)f_{m}(\beta) = {1\over Z(\beta)}\,e^{-\beta
E_{n}}\,\delta_{mn}\label{d6}
\end{equation}
Since $f_{n}$'s are ordinary numbers and eq. (\ref{d6}) is more like an
orthonormality condition, it is clear that we cannot satisfy this
condition (and, therefore, define a thermal vacuum with the right
properties) if we restrict ourselves to the original Hilbert space.

On the other hand, it is also clear from this analysis that if
$f_{n}$'s, somehow, behave like a state vector, then, the condition
in eq. (\ref{d6}) can be easily satisfied. In fact, let us introduce a
fictitious system identical to our original system and denote it by a
tilde system. The states in the combined Hilbert space of this doubled
system  would have the form
\[
|n,\tilde{m}\rangle = |n\rangle\otimes |\tilde{m}\rangle
\]
Let us assume that the thermal vacuum can be written as a linear
superposition of states in this doubled Hilbert space of the form
\begin{equation}
|0,\beta\rangle = \sum_{n} f_{n}(\beta) |n,\tilde{n}\rangle = \sum_{n}
 f_{n}(\beta) |n\rangle\otimes |\tilde{n}\rangle\label{d7}
\end{equation}
This would lead to
\begin{eqnarray}
\langle 0,\beta|{\cal O}|0,\beta\rangle & = & \sum_{n,m}
f_{n}^{*}(\beta)f_{m}(\beta)\,\langle n,\tilde{n}|{\cal
O}|m,\tilde{m}\rangle\nonumber\\
 & = & \sum_{n,m} f_{n}^{*}(\beta)f_{m}(\beta)\,\langle n|{\cal
O}|m\rangle\,\delta_{n,m}\nonumber\\
 & = & \sum_{n} f_{n}^{*}(\beta)f_{n}(\beta)\,\langle n|{\cal
O}|n\rangle\label{d8}
\end{eqnarray}

In deriving this result, we have used the fact that an operator of the
original system does not act on states of the tilde system and {\it
vice versa}. The result in eq. (\ref{d8}) is quite interesting because
it says that if we choose
\begin{equation}
f_{n}^{*}(\beta)f_{n}(\beta) = {e^{-\beta E_{n}}\over Z(\beta)}\qquad
{\rm or,}\;f_{n}(\beta)=f_{n}^{*}(\beta)={e^{-\beta E_{n}/2}\over
Z^{1/2}(\beta)}\label{d9}
\end{equation}
then, eq. (\ref{d8}) would, indeed, coincide with the ensemble average
in eq. (\ref{d3}). 

This analysis shows that it is possible to introduce a thermal vacuum
such that the ensemble average of any operator can be written as the
expectation value of the operator in the thermal vacuum. The price one
has to pay is that the Hilbert space needs to be doubled. The
advantage, on the other hand, lies in the fact that the description
would now involve both time and temperature (since we have not traded
time for temperature) and all the diagrammatic methods of zero
temperature field  theory can now be taken over directly.

\subsection*{Fermionic Oscillator}

To get a flavor for things in this formalism, let us analyze in some
detail the simple quantum mechanical system of the fermionic
oscillator. The Hamiltonian for the system is given by ($\hbar =1$)
\begin{equation}
H = \omega a^{\dagger}a\label{d10}
\end{equation}
Here, the fermionic creation and annihilation operators satisfy the
canonical anti-commutation relations
\begin{eqnarray}
\left[a , a^{\dagger}\right]_{+} & = & 1\nonumber\\
\left[a , a\right]_{+} & = & \left[a^{\dagger} ,
a^{\dagger}\right]_{+} = 0\label{d11}
\end{eqnarray}
In this case, the spectrum of the Hamiltonian is quite simple and the
Hilbert space is two dimensional with the basis states given by
$|0\rangle$ and $|1\rangle = a^{\dagger}|0\rangle$.

According to the general philosophy of thermo field dynamics, we are
supposed to introduce a fictitious tilde system which is identical to
our original system. Thus, we define
\begin{equation}
\widetilde{H} = \omega \tilde{a}^{\dagger}\tilde{a}\label{d12}
\end{equation}
with the anti-commutation relations
\begin{eqnarray}
\left[\tilde{a} , \tilde{a}^{\dagger}\right]_{+} & = & 1\nonumber\\
\left[\tilde{a} , \tilde{a}\right]_{+} & = & \left[\tilde{a}^{\dagger}
, \tilde{a}^{\dagger}\right]_{+} = 0\label{d13}
\end{eqnarray}
Furthermore, we assume that the creation and the annihilation
operators for the tilde and the non-tilde systems anti-commute.

The Hilbert space for the combined space is now four dimensional and,
following our earlier discussion, we choose the thermal vacuum to be
\begin{equation}
|0,\beta\rangle = f_{0}(\beta)|0\rangle\otimes |\tilde{0}\rangle +
 f_{1}(\beta)|1\rangle\otimes |\tilde{1}\rangle\label{d14}
\end{equation}
The normalization of the thermal vacuum gives
\begin{equation}
\langle 0,\beta|0,\beta\rangle = |f_{0}(\beta)|^{2} +
|f_{1}(\beta)|^{2} = 1\label{d15}
\end{equation}
while the expectation value of the number operator gives
\begin{equation}
\langle 0,\beta|N|0,\beta\rangle = \langle
0,\beta|a^{\dagger}a|0,\beta\rangle = |f_{1}(\beta)|^{2} = {1\over
e^{\beta\omega}+1}\label{d16}
\end{equation}
From these, we can obtain
\begin{equation}
f_{0}(\beta) = {1\over \sqrt{1+e^{-\beta\omega}}},\qquad f_{1}(\beta) =
{e^{-\beta\omega/2}\over \sqrt{1+e^{-\beta\omega}}}\label{d17}
\end{equation}
so that we can write
\begin{equation}
|0,\beta\rangle = {1\over
 \sqrt{1+e^{-\beta\omega}}}\left(|0,\tilde{0}\rangle +
 e^{-\beta\omega/2}\,|1,\tilde{1}\rangle\right)\label{d18}
\end{equation}   

To further understand the properties of this system, let us note that
we can define a Hermitian operator in this doubled space
\begin{equation}
G(\theta) = -i\theta(\beta)\left(\tilde{a}a -
a^{\dagger}\tilde{a}^{\dagger}\right)\label{d19}
\end{equation}
This would, in turn, lead to a formally unitary operator
\begin{equation}
U(\beta) = e^{-iG(\theta)}\label{d20}
\end{equation}
which would connect the thermal vacuum to the vacuum of the doubled
space, namely,
\begin{equation}
U(\beta)|0,\tilde{0}\rangle = \cos\theta(\beta) |0,\tilde{0}\rangle +
\sin\theta(\beta) |1,\tilde{1}\rangle = |0, \beta\rangle\label{d21}
\end{equation}
provided
\begin{equation}
\cos\theta(\beta) = f_{0}(\beta) = {1\over
\sqrt{1+e^{-\beta\omega}}},\qquad \sin\theta(\beta) = f_{1}(\beta) =
{e^{-\beta\omega/2}\over \sqrt{1+e^{-\beta\omega}}}\label{d22}
\end{equation}

The unitary operator would also induce a transformation on the
operators of the form
\begin{equation}
{\cal O}(\beta) = U(\beta){\cal O}U^{\dagger}(\beta)\label{d23}
\end{equation}
In particular, this would give
\begin{eqnarray}
a(\beta) & = & \cos\theta(\beta)\,a -
\sin\theta(\beta)\,\tilde{a}^{\dagger}\nonumber\\
\tilde{a}(\beta) & = & \cos\theta(\beta)\,\tilde{a} +
\sin\theta(\beta)\,a^{\dagger}\label{d24}
\end{eqnarray}
as well as their Hermitian conjugates. These operators would satisfy
the same anti-commutation relations as the original ones and we can
think of them as the thermal creation and annihilation
operators. Consequently, we can build up the thermal Hilbert space
starting from  $|0,\beta\rangle$ and the thermal creation operators.

In particular, it is trivial to check, using (\ref{d21}) that the
thermal vacuum satisfies
\begin{eqnarray}
a(\beta)|0,\beta\rangle & = & (\cos\theta(\beta)\,a -
\sin\theta(\beta)\,\tilde{a}^{\dagger})|0,\beta\rangle = 0\nonumber\\
\tilde{a}(\beta)|0,\beta\rangle & = & (\cos\theta(\beta)\,\tilde{a} +
\sin\theta(\beta)\,a^{\dagger})|0,\beta\rangle = 0\label{d25}
\end{eqnarray}
This is quite interesting for it says that annihilating a particle in
the thermal vacuum is equivalent to creating a tilde particle and {\it
vice versa}. Consequently, we can intuitively think of the tilde
particles as kind of hole states of the particles or particle states
of the heat bath. This gives a nice intuitive meaning to the doubling
of the degrees of freedom in thermo field dynamics. Namely, an isolated
system in thermal equilibrium really consists of two components -- the
original system and the heat bath.

We also note here that although the operator connecting the
thermal vacuum to the vacuum in the doubled space is formally unitary,
it is more like a Bogoliubov transformation. In more complicated
models with an infinite number of degrees of freedom (namely, in field
theories) such an operator takes us to a unitarily inequivalent
Hilbert space. Let us also note here, for future use, the simple formula
following from eq. (\ref{d24}) that
\begin{equation} 
\left(\begin{array}{c}
       a(\beta)\\
       \tilde{a}^{\dagger}(\beta)
       \end{array}\right) = \widetilde{U}(\beta)\left(\begin{array}{c}
       a\\
       \tilde{a}^{\dagger}
       \end{array}\right)\label{d26}
\end{equation}
where
\begin{equation}
\widetilde{U}(\beta) = \left(\begin{array}{rr}
                        \cos\theta(\beta) & -\sin\theta(\beta)\\
                        \sin\theta(\beta) & \cos\theta(\beta)
                        \end{array}\right)\label{d27}
\end{equation}

Finally, let us conclude the discussion of this example by noting that
the  states in the
thermal Hilbert space are eigenstates of neither $H$ nor
$\widetilde{H}$. Rather they are the eigenstates of the operator
\begin{equation}
\widehat{H} = H - \widetilde{H}\label{d28}
\end{equation}
Furthermore, this combination of the Hamiltonians is also invariant
under the unitary transformation of (\ref{d21}). This is, indeed, the
Hamiltonian that governs the dynamics of the combined system.

\subsection*{Bosonic Oscillator}

The analysis for the case of the bosonic oscillator is quite analogous
to the discussion of the fermionic oscillator. Therefore, without
going into too much detail, let us summarize the results. First, the
Hamiltonian for the system is given by
\begin{equation}
H = \omega a^{\dagger} a
\end{equation}
much like the fermionic oscillator. However, the creation and
annihilation operators satisfy canonical commutation relations of the
form
\begin{eqnarray}
\left[a , a^{\dagger}\right] & = & 1\nonumber\\
\left[a , a\right] & = & \left[a^{\dagger} , a^{\dagger}\right] =
0\label{d29}
\end{eqnarray}
The Hilbert space for the bosonic oscillator is infinite dimensional
with the energy eigenstates given by
\begin{equation}
H|n\rangle = n\omega\,|n\rangle,\qquad n=0,1,2,\cdots\label{d30}
\end{equation}

According to the general discussions of thermo field dynamics, we
introduce an identical, but fictitious tilde system with the
Hamiltonian
\begin{equation}
\widetilde{H} = \omega \tilde{a}^{\dagger}\tilde{a}\label{d31}
\end{equation}
The tilde creation and annihilation operators are expected to satisfy
commutation relations analogous to (\ref{d29}). Furthermore, the tilde
operators are supposed to commute with the original operators of the
theory.

Following the discussion of the earlier section, we can determine the
thermal vacuum state in this case to be
\begin{equation}
|0,\beta\rangle =
 (1-e^{-\beta\omega})^{1/2}\,\sum_{n=0}^{\infty}\,e^{-n\beta\omega/2}\,|n,
\tilde{n}\rangle\label{d32} 
\end{equation}
As in the fermionic oscillator, we can introduce the Hermitian
operator
\begin{equation}
G(\theta) = -i\theta(\beta)\left(\tilde{a}a -
a^{\dagger}\tilde{a}^{\dagger}\right)\label{d33}
\end{equation}
and the unitary operator
\begin{equation}
U(\beta) = e^{-iG(\theta)}\label{d34}
\end{equation}
Then, it is straightforward to check and see that the unitary operator
connects the thermal vacuum to the vacuum of the doubled space
provided
\begin{equation}
\cosh\theta(\beta) = {1\over \sqrt{1-e^{-\beta\omega}}}\qquad
\sinh\theta(\beta) = {e^{-\beta\omega/2}\over
\sqrt{1-e^{-\beta\omega}}}\label{d35}
\end{equation}

The unitary operator induces a transformation of the operators of the
form
\begin{equation}
{\cal O}(\beta) = U(\beta){\cal O}U^{\dagger}(\beta)\label{d36}
\end{equation}
leading to
\begin{eqnarray*}
a(\beta) & = & \cosh\theta(\beta)\,a -
\sinh\theta(\beta)\,\tilde{a}^{\dagger}\\
\tilde{a}(\beta) & = & \cosh\theta(\beta)\,\tilde{a} -
\sinh\theta(\beta)\,a^{\dagger}
\end{eqnarray*}
and similarly for the Hermitian conjugates. As we have seen in the
last section, these can be thought of as the creation and annihilation
operators for the thermal Hilbert space. In particular, the thermal
vacuum is easily seen to satisfy
\begin{eqnarray}
a(\beta)|0,\beta\rangle & = & (\cosh\theta(\beta)\,a -
\sinh\theta(\beta)\,\tilde{a}^{\dagger})|0,\beta\rangle = 0\nonumber\\
\tilde{a}(\beta)|0,\beta\rangle & = & (\cosh\theta(\beta)\,\tilde{a} -
\sinh\theta(\beta)\,a^{\dagger})|0,\beta\rangle = 0\label{d37}
\end{eqnarray}
This, again, reinforces the intuitive picture of doubling in
thermo field dynamics. Let us also note here, for future use, the
simple formula following from (\ref{d36})
\begin{equation} 
\left(\begin{array}{c}
       a(\beta)\\
       \tilde{a}^{\dagger}(\beta)
       \end{array}\right) = \widetilde{U}(\beta)\left(\begin{array}{c}
       a\\
       \tilde{a}^{\dagger}
       \end{array}\right)\label{d38}
\end{equation}
where
\begin{equation}
\widetilde{U}(\beta) = \left(\begin{array}{rr}
                        \cosh\theta(\beta) & -\sinh\theta(\beta)\\
                        -\sinh\theta(\beta) & \cosh\theta(\beta)
                        \end{array}\right)\label{d39}
\end{equation}

\subsection*{Field Theory}

The extension of these results to a field theory is quite
straightforward once we keep in mind that, at the free level, a
quantum field theory is simply an infinite  collection of oscillators
with frequencies dependent on the momentum of the mode. Consequently,
the thermal vacuum, in this case, would be connected to the vacuum of
the doubled space as
\begin{equation}
|0,\beta\rangle = U(\beta)|0,\tilde{0}\rangle =
 e^{-iG(\theta)}\,|0,\tilde{0}\rangle\label{d40}
\end{equation}
where
\begin{equation}
G(\theta) =
-i\,\sum_{\vec{k}}\,\theta_{\vec{k}}(\beta)\left(\tilde{a}_{\vec{k}}a_{\vec{k}}
- a_{\vec{k}}^{\dagger}\tilde{a}_{\vec{k}}^{\dagger}\right)\label{d41}
\end{equation}
with, say, for bosons,
\begin{equation}
\cosh\theta_{\vec{k}}(\beta) = {1\over
\sqrt{1-e^{-\beta\omega_{k}}}}\qquad \sinh\theta_{\vec{k}}(\beta) =
{e^{-\beta\omega_{k}/2}\over \sqrt{1-e^{-\beta\omega_{k}}}}\label{d42}
\end{equation}
Here, for a relativistic theory, we have
\[
\omega_{k} = \sqrt{\vec{k}^{2} + m^{2}}
\]

Let us next note that, at zero temperature, the original fields are
decoupled from the tilde fields. Thus, if we were to define a doublet
of fields (real scalar field) as in eq. (\ref{d38})
\begin{equation}
\Phi = \left(\begin{array}{c}
              \phi\\
              \tilde{\phi}
              \end{array}\right)\label{d43}
\end{equation}
then, at zero temperature the propagator is defined to be (This is not
to be confused with the generator of the Bogoliubov transformations in
eqs. (\ref{d19}), (\ref{d33}) and (\ref{d41}))
\[
iG(x-y) = \langle 0,\tilde{0}|T(\Phi(x)\Phi(y))|0,\tilde{0}\rangle
\]
which has the momentum space representation
\begin{equation}
G(k) = \left(\begin{array}{cc}
             {1\over k^{2}-m^{2}+i\epsilon} & 0\\
              0 & -{1\over k^{2}-m^{2}-i\epsilon}
              \end{array}\right)\label{d44}
\end{equation}
Given this, the finite temperature propagator can be determined to be
\begin{eqnarray}
iG_{\beta}(x-y) & = & \langle
0,\beta|T(\Phi(x)\Phi(y))|0,\beta\rangle\nonumber\\
 & = & \langle
0,\tilde{0}|U^{\dagger}(\beta)T(\Phi(x)\Phi(y))U(\beta)|0,\tilde{0}\rangle
\label{d45}
\end{eqnarray}
Using now the generalization of eqs. (\ref{d36},\ref{d38},\ref{d39}), the
momentum representation for the propagator can be determined to be
\begin{eqnarray}
G_{\beta}(k) & = &
\widetilde{U}(-\theta_{\vec{k}})G(k)\widetilde{U}^{T}(-\theta_{\vec{k}})
\nonumber\\
& = & \left(\begin{array}{cc}
             {1\over k^{2}-m^{2}+i\epsilon} & 0\\
              0 & -{1\over k^{2}-m^{2}-i\epsilon}
              \end{array}\right)
      -2i\pi n_{B}(|k^{0}|)\delta(k^{2}-m^{2})\left(\begin{array}{rr}
                                                 1 &
e^{\beta|k^{0}|/2}\\
e^{\beta|k^{0}|/2} & 1
\end{array}\right)\label{d46}
\end{eqnarray}

There are several things to note from the structure of the propagator
which are quite  general for a real time formalism. First, the propagator is a
$2\times 2$ matrix, a consequence of the doubling of the degrees of
freedom. Second, the propagator is a sum of two parts -- one
representing the zero temperature part and the other representing the
true temperature dependent corrections. The propagator is still the
Greens function for the free operator of the theory, but corresponding
to different boundary conditions (remember the KMS condition in
eq. (\ref{c3})). While the zero temperature part of the propagator
corresponds, as usual, to the exchange of a virtual particle, the
temperature dependent part represents an on-shell contribution
(because of the delta function). In fact, the intuitive meaning of the
temperature dependent correction is quite clear. In a hot medium,
there is a distribution of real particles and the temperature
dependent part merely represents the possibility that a particle, in
addition to having virtual exchanges, can also emit or absorb a real
particle of the medium. 

Since the temperature dependent part of the propagator is on-shell, it
is clear that there can be no new ultraviolet divergence generated at
finite temperature. All the counter terms needed to renormalize the
theory at zero temperature would be sufficient for studies at finite
temperature as well. (Of course, the infrared behavior is another
story. Infrared divergence, in a field theory, becomes much more
severe at finite temperature, a topic that I will not get into.) There
is an alternate way to visualize this. At finite temperature the
distribution of the real particles is Boltzmann suppressed as we go
up in energy and, consequently, thermal corrections corresponding to
infinite  energy cannot arise. 

Once, we have the propagator, we can venture to do a diagrammatic
calculation in this formalism. The only things missing are the
interaction vertices of the theory. There is a well defined procedure [26]
(called the tilde conjugation rule) to construct the complete
Lagrangian from which to construct the vertices. Very simply, it
corresponds to what we have noted earlier, namely, the dynamical
Hamiltonian (and, similarly, the Lagrangian) is as given in eq.
(\ref{d28}). It is simply the difference between the original and the
tilde Hamiltonians. Thus, we see that the complete theory would
contain two kinds of vertices -- one for the original fields while the
second for the tilde fields. The vertices for the tilde fields will
have a relative negative sign corresponding to the original
vertices. Given the vertices and the propagator, it is now
straightforward to carry out any diagrammatic calculation to any
order. Let me emphasize here that although, at the tree level, there
is no vertex containing both the original and the tilde fields, such
vertices would be generated at higher loops because of the nontrivial
matrix structure of the propagator.

Thermo Field dynamics is a real time formalism. But, more than that, it
is really an operator formalism and hence very well suited to study
various operator questions such as the structure of the thermal
vacuum, the theorems on symmetry breaking etc. It can also be given a
path integral representation and corresponds to choosing a specific
time contour in the complex $t$ plane [27, 10] (remember that the imaginary
time formalism also corresponds to choosing a specific time contour,
namely, along the imaginary time axis) and I will come back to this
question in the next section. However, once again from the philosophy
of thermo field dynamics, it is clear that, it is a natural formalism
to describe equilibrium phenomena where quantities depend on both time
and temperature. While there are several attempts to generalize this
to include non-equilibrium phenomena, there does not yet exist a complete
description.

\section{Closed Time Path Formalism}

The closed time path formalism is also a real time formalism which was
formulated much earlier than thermo field dynamics within the context
of non-equilibrium phenomena [2]. The two formalisms are, in some sense,
complementary to each other although the closed time path formalism
can describe both equilibrium and non-equilibrium phenomena with equal
ease.

The basic idea behind the closed time path formalism [10] is the fact that
when a quantum mechanical system is in a mixed state, as is the case
in the presence of a heat bath, the system can be naturally described
in terms of a density matrix defined, in the Schr\"{o}dinger picture,  as
\begin{equation}
\rho(t) = \sum_{n} p_{n}\,|\psi_{n}(t)\rangle \langle
\psi_{n}(t)|\label{e1}
\end{equation}
Here, $p_{n}$ represents the probability for finding the quantum
mechanical system in the state $|\psi_{n}(t)\rangle$ and, for
simplicity, we have assumed the quantum mechanical states to form a
discrete set. It is $p_{n}$ which contains information regarding the
surrounding which is hard to determine, but, being a probability, it
satisfies
\[
\sum_{n} p_{n} = 1
\]

Given the density matrix,  the ensemble average of any
operator can be calculated in the Schr\"{o}dinger picture as
\begin{equation}
\langle {\cal O}\rangle (t) = \sum_{n} p_{n}\,\langle
\psi_{n}(t)|{\cal O}|\psi_{n}(t)\rangle = {\rm Tr}\,\rho(t)\,{\cal
O}\label{e2} 
\end{equation}
The ensemble average, in this case, naturally develops a time
dependence from the time dependence of the density matrix. In this
formalism, we can naturally define an entropy as
\[
S = - \sum_{n} p_{n}\,\ln p_{n}
\]
which is by definition positive semi-definite and measures the order
(or lack of it) in an ensemble.

The state vectors satisfy the Schr\"{o}dinger equation ($\hbar = 1$)
\[
i\,{\partial|\psi(t)\rangle\over \partial t} = H|\psi(t)\rangle
\]
From this, we can determine the time evolution of the density matrix
which turns out to be the Liouville equation
\begin{equation}
i\,{\partial\rho(t)\over \partial t} = \left[H ,
\rho(t)\right]\label{e3}
\end{equation}
In deriving this, we have assumed that the probabilities do not change
with time (appreciably) implying that entropy remains constant during
such an evolution. The reason for this assumption is our lack of
knowledge about the time evolution of the surrounding such as the heat
bath. On the other hand, adiabatic evolutions do arise frequently in
physical systems and, consequently, we would continue with this
assumption.

Let us note that eq. (\ref{e3}) has a simple solution of the form
\begin{equation}
\rho(t) = U(t,0)\rho(0)U^{\dagger}(t,0) =
U(t,0)\rho(0)U(0,t)\label{e4}
\end{equation}
where the time evolution operator has the general form
\begin{equation}
U(t,t') = T\left(e^{-i\int_{t'}^{t} dt''\,H(t'')}\right)\label{e5}
\end{equation}
Furthermore, it satisfies the semi-group properties
\begin{eqnarray}
U(t_{1},t_{2})U(t_{2},t_{1}) & = & 1\nonumber\\
U(t_{1},t_{2})U(t_{2},t_{3}) & = & U(t_{1},t_{3}) \qquad {\rm
for}\;t_{1}>t_{2}>t_{3}\label{e6}
\end{eqnarray}
In particular, let us note that if the Hamiltonian is time
independent, eq. (\ref{e4}) takes the simple form
\[
\rho(t) = e^{-iHt}\rho(0) e^{iHt}
\]
and, furthermore, if the Hamiltonian commutes with $\rho(0)$, the
density  matrix
would be time independent, describing a system in equilibrium. This
would be true, for example, if the states in eq. (\ref{e1}) are
stationary states. This is also true if the probabilities have a
Boltzmann distribution in which case, we refer to the system as being
in thermal equilibrium. However, we will not restrict to any such
special case allowing for the formalism to accommodate both equilibrium
and non-equilibrium phenomena.

Keeping in mind the fact that we are ultimately interested in a thermal
ensemble, let us choose
\begin{equation}
\rho(0) = {e^{-\beta H_{i}}\over {\rm Tr}\,e^{-\beta H_{i}}}\label{e7}
\end{equation}
for some $H_{i}$. Since the density matrix is a positive Hermitian
matrix with unit
trace, mathematically, this is allowed. But, more important is the
physical reason behind such a choice. Namely, we can think of the
dynamical Hamiltonian of our system as
\begin{equation}
H(t) = \left\{\begin{array}{cll}
              H_{i} & {\rm for} & {\rm Re}t\leq 0\\
             {\cal H}(t) & {\rm for} & {\rm Re}t\geq 0
             \end{array}\right.\label{e8}
\end{equation}
This would correspond to the fact that we prepare our system in a
equilibrium state at temperature ${1\over\beta}$ for negative times and let
the system evolve, for positive times, with the true Hamiltonian
${\cal H}$ which may be time dependent. If ${\cal H}(t) = H_{i}$,
then, the system will evolve in equilibrium and not otherwise.

With eq. (\ref{e8}) in mind, we note that we can write
\begin{equation}
\rho(0) = {U(T-i\beta,T)\over {\rm Tr}\,U(T-i\beta,T)}\label{e9}
\end{equation}
where $T$ is assumed to be a large negative time (and not the
temperature) and $T\rightarrow -\infty$ at the end. Using the
semi-group properties of the time evolution operator, it is easy to
see that the ensemble average of any operator can now be represented
as
\begin{eqnarray}
\langle {\cal O}\rangle_{\beta} & = & {\rm Tr}\,\rho(t)\,{\cal O}\nonumber\\
 & = & {{\rm Tr}\,U(t,0)\,U(T-i\beta,T)\,U(0,t)\,{\cal O}\over {\rm
 Tr}\,U(T-i\beta,T)}\nonumber\\
 & = & {{\rm Tr}\,U(T-i\beta,T)\,U(T,T')\,U(T',t)\,{\cal
 O}\,U(t,T)\over {\rm Tr}\,U(T-i\beta,T)\,U(T,T')\,U(T',T)}\label{e10}
\end{eqnarray}
where we have introduced a large positive time $T'$ and assume that
$T'\rightarrow \infty$ at the end. This gives a nice representation to
the ensemble average of any operator. Namely, we let the system evolve
from a large negative time $T$ to $t$ where the appropriate operator
${\cal O}$ is inserted. The system then, evolves from $t$ to a large
positive time $T'$ and back from $T'$ to $T$ and then, continues evolving along
the imaginary branch from $T$ to $T-i\beta$. Since the matrix elements
of the time evolution operator can be given a path integral
representation, it is clear that the ensemble average of any operator
can also be given a path integral representation in this formalism
corresponding to the specific  contour in the complex time plane
as described above. Although the specific contour has three branches --
one along the real axis increasing with time, the second also along
the real axis decreasing with time and the third along the negative
imaginary axis --  in the limit $T\rightarrow -\infty$ and
$T'\rightarrow\infty$, it can be shown that the third branch gets
decoupled from the other two (the factors in the propagators
connecting such branches are asymptotically damped). Consequently, in
this limit, we are effectively dealing with two branches leading to
the name \lq\lq closed time path formalism'' [12]. In this contour, then,
the time integration has to be thought of as
\begin{equation}
\int_{c} dt = \int_{-\infty}^{\infty} dt_{+} - \int_{-\infty}^{\infty}
dt_{-}\label{e11}
\end{equation}
where the relative negative sign arises because time is decreasing in
the second branch of the time contour.

The doubling of the degrees of freedom, in this formalism, is now
clear. To have a path integral description, we must specify the fields
on both the branches of the contour. Or, equivalently, we can use just
the positive branch and double the field degrees of freedom. Namely,
corresponding to every original field, say $\phi_{+}$, we must
introduce a second field $\phi_{-}$ and remember that the action for
the $\phi_{-}$ fields must have a relative negative sign arising from
eq. (\ref{e11}), namely, that time is decreasing along the second
branch.

\subsection*{Scalar Field Theory}

Just as an example, let us study next the self-interacting scalar field
theory in some detail. The Lagrangian density is the same as in
eq. (\ref{c21}), but following the earlier discussion, we should take
the complete Lagrangian density for the system to be
\begin{equation}
{\cal L} = {\cal L}(\phi_{+}) - {\cal L}(\phi_{-})\label{e12}
\end{equation}
where
\begin{equation}
{\cal L}(\phi) = {1\over 2}\partial_{\mu}\phi\partial^{\mu}\phi -
{m^{2}\over 2}\phi^{2} - {\lambda\over 4!}\phi^{4}\qquad \lambda>0\label{e13}
\end{equation}
The Feynman propagator can again be determined for this theory and
would have a $2\times 2$ matrix structure because of the doubling of
the field degrees of freedom. It can be determined subject to
compatibility with the KMS conditions and has the form in the momentum
space 
\begin{equation}
G(k) = \left(\begin{array}{cc}
G_{++}(k) & G_{+-}(k)\\
G_{-+}(k) & G_{--}(k)
\end{array}\right)\label{e14}
\end{equation}
with
\begin{eqnarray}
G_{++}(k) & = & {1\over k^{2}-m^{2}+i\epsilon} - 2i\pi
n_{B}(|k^{0}|)\delta(k^{2}-m^{2})\nonumber\\
G_{+-}(k) & = & -2i\pi\left(\theta(-k^{0})+n_{B}(|k^{0}|)\right)
\delta(k^{2}-m^{2})\nonumber\\
G_{-+}(k) & = & -2i\pi\left(\theta(k^{0})+n_{B}(|k^{0}|)\right)
\delta(k^{2}-m^{2})\nonumber\\
G_{--}(k) & = & {1\over k^{2}-m^{2}-i\epsilon} - 2i\pi n_{B}(|k^{0}|)
\delta(k^{2}-m^{2})\label{e15}
\end{eqnarray}

There are several things to note from the structure of this
propagator. First, as in the case of the propagator in thermo field
dynamics, here, too, we see that the propagator naturally is a sum of
two parts -- the temperature independent part and the temperature
dependent part. But, more interestingly, here the propagator has the
simplification that the temperature dependent part of every component
is the same which leads to various simplifications in actual studies
of thermal quantities. Furthermore, not all the components of the
propagator  are independent. In fact, it is easily seen that (this can
be  traced back to their definition)
\[
G_{++}(k) + G_{--}(k) = G_{+-}(k) + G_{-+}(k)
\]
These are known as the causal propagators of the theory and are useful
in diagrammatic evaluation. There is, of course, another kind of
propagator, conventionally known as the physical propagators and is
defined as
\begin{equation}
\hat{G}(k) = \left(\begin{array}{cc}
0 & G_{A}(k)\\
G_{R}(k) & G_{C}(k)
\end{array}\right)\label{e16}
\end{equation}
where $G_{A}$, $G_{R}$ and $G_{C}$ are known as the advanced, retarded
and the correlated Greens functions. These are quite useful in the
study of various phenomena such as the linear response theory. The
important thing to observe is that the causal and the physical
propagators are connected through a unitary transformation
\begin{equation}
\hat{G}(k) = Q\,G\,Q^{\dagger}\label{e17}
\end{equation}
where
\begin{equation}
Q = {1\over \sqrt{2}}\left(\begin{array}{rr}
1 & -1\\
1 & 1
\end{array}\right)\label{e18}
\end{equation}
It can be determined from this that, at the tree level,
\begin{eqnarray}
G_{A}(k) & = & {1\over k^{2}-m^{2}-i\epsilon k^{0}}\nonumber\\
G_{R}(k) & = & {1\over k^{2}-m^{2}+i\epsilon k^{0}}\nonumber\\
G_{C}(k) & = & -2i\pi\left(1 +
2n_{B}(|k^{0}|)\right)\delta(k^{2}-m^{2})\label{e19}
\end{eqnarray}
as they should be.

The diagrammatic calculations can now be easily understood in this
formalism. The vertices can be read out from the Lagrangian density in
eq. (\ref{e12}). There are two kinds of vertices, one for the original
fields, $\phi_{+}$, and the other for the doubled fields,
$\phi_{-}$. The vertices for the $\phi_{-}$ fields are the same as
those for the $\phi_{+}$ fields except for a relative sign. With the
vertices and the causal propagators, one can now carry out the
calculation of any observable to any order in perturbation theory. As
before, we note that, although there is no coupling between the
$\phi_{+}$ and $\phi_{-}$ fields at the tree level, higher order
corrections would, in general, couple them.

As an example, let us calculate the one loop mass correction in this
theory. There will be two such diagrams to calculate -- one for the
$\phi_{+}$ field and the other for the $\phi_{-}$ field. The mass
correction for the $\phi_{+}$ field is readily seen to be
\begin{eqnarray}
-i\Delta m_{+}^{2} & = & {(-i\lambda)\over 2}\int {d^{4}k\over
 (2\pi)^{4}}\,iG_{++}(k)\nonumber\\
 & = & {\lambda\over 2}\int {d^{4}k\over (2\pi)^{4}}\left({1\over
 k^{2}-m^{2}+i\epsilon}-2i\pi
 n_{B}(|k^{0}|)\delta(k^{2}-m^{2})\right)\nonumber\\
 & = & -i(\Delta m_{0}^{2} + \Delta m_{\beta}^{2})\label{e20}
\end{eqnarray}
where it is easily seen that the temperature independent part has the
form 
\begin{equation}
\Delta m_{0}^{2} = {\lambda\over 4}\int {d^{3}k\over
(2\pi)^{3}}\,{1\over \omega_{k}}\label{e21}
\end{equation}
while the temperature dependent part is given by
\begin{equation}
\Delta m_{\beta}^{2} = {\lambda\over 2}\int {d^{3}k\over
(2\pi)^{3}}\,{n_{B}(\omega_{k})\over \omega_{k}} = {\lambda\over
2}\int {d^{3}k\over (2\pi)^{3}}\,{1\over \omega_{k}}\,{1\over
e^{\beta\omega_{k}} -1}\label{e22}
\end{equation}
These can be compared with the corresponding terms in
eq. (\ref{c26}). We can also calculate the mass correction for the
$\phi_{-}$ field. With a little bit of analysis, it is seen that
\[
\Delta m_{-}^{2} = \Delta m_{+}^{2}
\]

\section{Feynman Parameterization}

So far, we have described the various formalisms that can be used to
do calculations at finite temperature. However, actual calculations
lead to many subtle, but interesting features of theories at finite
temperature. One immediate and obvious feature, of course, is that finite
temperature effects break Lorentz invariance. Namely, in studying a
system at finite temperature, one has to go to a specific frame where
the heat bath is at rest and, consequently, Lorentz invariance will no
longer be manifest. This is, of course, already manifest at the level
of propagators. For example,  the structure of the propagators in
eqs. (\ref{d46}) or (\ref{e15}) clearly displays a Lorentz
non-invariant structure. The consequence of this is that an amplitude
calculated at finite temperature, say for example, the self-energy
$\Pi(p^{0},\vec{p})$ depends on the external energy and momentum
independently. In fact, the self-energy becomes a non-analytic
function of these two variables at the origin and two different ways of
approaching the origin in this space leads to distinct  plasmon and 
screening masses [23]. Thus, such non-analyticities are quite physical and
their origin can be traced back to the fact that, at finite
temperature, there are new channels of reactions possible leading to
new branch cuts which give rise to such discontinuities [23, 10]. (To be
absolutely fair, it is worth noting that statistical mechanics can be
formulated in a covariant way. In such a case, one finds that there
is a larger number of Lorentz invariant variables that can be defined
on which amplitudes can depend. The non-analyticity in $p^{0}$ and
$\vec{p}$ can then be translated to a non-analyticity in these new, Lorentz
invariant variables [23].)

There are, of course, some other kinds of subtlety that arise which
influence the calculations directly at finite temperature. We will
discuss one such subtlety in this section. Let us note that a
particularly useful formula in the evaluation of amplitudes at zero
temperature is the Feynman combination formula given by
\begin{equation}
{1\over A+i\epsilon}{1\over B+i\epsilon} = \int_{0}^{1}
{dx\over [x(A+i\epsilon) +
(1-x)(B+i\epsilon)]^{2}}\label{f1}
\end{equation}
This can be directly checked by evaluating the $x$ integral on the
right hand side.

This formula is extremely useful and works at zero temperature mainly
because the Feynman propagators have the same analytic structure,
namely, they have the same \lq\lq $i\epsilon$'' dependence. In
contrast, we note that the finite temperature propagators contain
delta functions (see eqs. (\ref{d46}) and (\ref{e15})) and recalling
that
\[
\delta(x) = \lim_{\epsilon\rightarrow 0^{+}} {1\over 2i\pi}
\left({1\over x - i\epsilon} - {1\over x + i\epsilon}\right)
\]
we recognize that, at finite temperature, the propagators no longer
have the same \lq\lq $i\epsilon$'' dependence. Consequently, in
evaluating Feynman amplitudes at finite temperature, we have to
combine denominators which do not necessarily have the same \lq\lq
$i\epsilon$'' dependence. Keeping this in mind, let us examine the
combination of two different denominators with arbitrary analytic
dependence. 

Without loss of generality, let us choose $\alpha,\beta = \pm 1$ and
note that
\begin{eqnarray}
\int_{0}^{1} {dx\over
\left[x(A+i\alpha\epsilon)+(1-x)(B+i\beta\epsilon)\right]^{2}} & = &
-{1\over (A-B)+i(\alpha-\beta)\epsilon}\left.{1\over
x(A+i\alpha\epsilon)+(1-x)(B+i\beta\epsilon)}\right|_{0}^{1}\nonumber\\
 & = & {1\over A+i\alpha\epsilon}\,{1\over B+i\beta\epsilon}\label{f2}
\end{eqnarray}
This is, of course, the standard Feynman combination formula. However,
let us note that this will not hold if  $0<x_{0}<1$ such that
\begin{equation}
x_{0} = {\beta\over \beta - \alpha},\qquad \beta\,A =
\alpha\,B\label{f3}
\end{equation}
because, in such a case, the integrand will have a singularity on the
real $x$-axis inside the interval of integration. In this case, we
have
\begin{eqnarray}
\int_{0}^{1}
{dx\over\left[x(A+i\alpha\epsilon)+(1-x)(B+i\beta\epsilon)\right]^{2}}
& = &
\int_{0}^{1} {dx\over
\left[x(A-B+i(\alpha-\beta)\epsilon)+B+i\beta\epsilon\right]^{2}}
\nonumber\\
& = &\!\!\! \lim_{\eta\rightarrow 0}\!\!\left(\!\!\int_{0}^{{\beta\over
\beta-\alpha}-\eta}+\!\int_{{\beta\over
\beta-\alpha}+\eta}^{1}\!\right)\!\! {dx\over 
\left[x(A-B+i(\alpha-\beta)\epsilon)+B+i\beta\epsilon\right]^{2}}
\nonumber\\
 & = & {1\over A+i\alpha\epsilon}\,{1\over B+i\beta\epsilon} - 2i\pi
{(\alpha-\beta)\delta(\beta\,A-\alpha\,B)\over A-B +
i(\alpha-\beta)\epsilon}\label{f4}
\end{eqnarray}
In other words, when the parameters of the integrand satisfy
eq. (\ref{f3}), the Feynman combination formula of eq. (\ref{f1}) will
modify and the general formula follows from eq. (\ref{f4}) to be [13]
\begin{equation}
{1\over A+i\alpha\epsilon}\,{1\over B+i\beta\epsilon} = \int_{0}^{1}
{dx\over \left[x(A+i\alpha\epsilon) +
(1-x)(B+i\beta\epsilon)\right]^{2}} + 2i\pi
{(\alpha-\beta)\delta(\beta A-\alpha B)\over
A-B+i(\alpha-\beta)\epsilon}\label{f5}
\end{equation}
Note that, condition (\ref{f3}) can only be satisfied (with
$0<x_{0}<1$) if $\alpha$ and $\beta$ are of opposite sign. Indeed, let
us note from eq. (\ref{f5}) that the second term vanishes when
$\alpha=\beta=1$ as is the case at zero temperature. Namely, when
propagators with identical \lq\lq $i\epsilon$'' dependence are combined,
the standard combination formula of eq. (\ref{f1}) holds. However, if
denominators with opposite \lq\lq $i\epsilon$'' dependence are
combined, the correct combination formula involves a second term. This
is quite crucial at finite temperature and without this second term,
one ends up with a wrong result as was discovered in finite
temperature calculations the hard way [28].

\section{Large Gauge Invariance}

Gauge theories are beautiful theories which describe physical forces
in a natural manner and because of their rich structure, the study of
gauge theories at finite temperature  is quite interesting in itself.
However, to avoid getting into technicalities, we will
not discuss the intricacies of such theories either at zero
temperature or at finite temperature. Rather, we will discuss a simple
quantum mechanical model, in this section, to bring out some of the
new features that temperature brings into such theories -- features
which are very different from what we expect at zero temperature.

To motivate, let us note that gauge invariance is realized as an
internal symmetry in quantum mechanical systems. Consequently, we do
not expect a macroscopic external surrounding such as a heat bath to
modify gauge invariance. This is more or less what is also found by
explicit computations at finite temperature, namely, that gauge
invariance and Ward identities continue to hold even at finite
temperature [29]. This is certainly the case when one is talking about
small gauge transformations for which the parameters of transformation
vanish at infinity.

However, there is a second class of gauge invariance, commonly known
as large gauge invariance where the parameters do not vanish at
infinity and this brings in some new topological character to physical
theories. For example, let us consider a $2+1$ dimensional
Chern-Simons theory of the form
\begin{eqnarray}
{\cal L} & = & M{\cal L}_{\rm CS} + {\cal L}_{\rm fermion}\nonumber\\
 & = &
 M\epsilon^{\mu\nu\lambda}\,{\rm tr}\,A_{\mu}(\partial_{\nu}A_{\lambda}-{2\over
 3}A_{\nu}A_{\lambda}) +
 \overline{\psi}(\gamma^{\mu}(i\partial_{\mu}-gA_{\mu})-m)\psi\label{g1}
\end{eqnarray}
where $M$ is a mass parameter, $A_{\mu}$  a matrix valued non-Abelian
gauge  field and \lq\lq tr'' stands for the matrix trace. The first
term, on the right hand side, is known as the Chern-Simons term which
exists only in odd space-time dimensions. We can, of course, also add
a Maxwell like term to the Lagrangian and, in that case, the
Chern-Simons term behaves like a mass term for the gauge
field. Consequently, such a term is also known as a topological mass
term [30](topological because it does not involve the metric). For
simplicity of discussion, however, we will not include a Maxwell like
term to the Lagrangian.

Under a gauge transformation of the form
\begin{eqnarray}
\psi & \rightarrow & U^{-1}\,\psi\nonumber\\
A_{\mu} & \rightarrow & U^{-1}\,A_{\mu}\,U - {i\over
g}\,U^{-1}\,\partial_{\mu}U\label{g2}
\end{eqnarray}
it is straightforward to check that the  action in eq. (\ref{g1}) is not
invariant, rather it changes as 
\begin{equation}
S = \int d^{3}x\,{\cal L} \rightarrow S + {4\pi M\over g^{2}}\,2i\pi
W\label{g3}
\end{equation}
where 
\begin{equation}
W = {1\over 24\pi^{2}}\int d^{3}x\,\epsilon^{\mu\nu\lambda}\,{\rm
tr}\partial_{\mu}UU^{-1}\partial_{\nu}UU^{-1}\partial_{\lambda}UU^{-1}
\label{g4}
\end{equation}
is known as the winding number. It is a topological quantity which is an
integer (Basically, the fermion Lagrangian density is invariant under
the gauge transformations, but the Chern-Simons term changes by a
total divergence which does not vanish if the gauge transformations do
not vanish at infinity. Consequently, the winding number counts the
number of times the gauge transformations  wrap around the
sphere.). For small gauge transformations, the winding number vanishes
since the gauge transformations vanish at infinity.

Let us note from eq. (\ref{g3}) that even though the action is not
invariant under a large gauge transformation, if $M$ is quantized in
units of ${g^{2}\over 4\pi^{2}}$, the change in the action would be a
multiple of $2i\pi$ and, consequently, the path integral would be
invariant under a large gauge transformation. Thus, we have the
constraint coming from the consistency of the theory that the
coefficient of the Chern-Simons term must be quantized. We have
derived this conclusion from an analysis of the tree level behavior of
the theory and we have to worry if the quantum corrections can change
the behavior of the theory. At zero temperature, an analysis of the
quantum corrections shows that the theory continues to be well defined
with the tree level quantization of the Chern-Simons coefficient
provided the number of fermion flavors is even. The even number of
fermion flavors is also necessary for a global anomaly of the theory to
vanish and so, everything is well understood at zero temperature.

At finite temperature, however, the situation appears to change
drastically. Namely, the fermions induce a temperature dependent
Chern-Simons term effectively making [31]
\begin{equation}
M \rightarrow M -{g^{2}\over 4\pi}\,{mN_{f}\over 2|m|}\,\tanh
{\beta|m|\over 2}\label{g5}
\end{equation}
Here, $N_{f}$ is the number of fermion flavors and this shows that, at
zero temperature ($\beta\rightarrow\infty$), $M$ changes by an integer
(in units of $g^{2}/4\pi$) for an even number of flavors. However, at
finite temperature, this becomes a continuous function of temperature
and, consequently, it is clear that it can no longer be an integer for
arbitrary values of the temperature. It seem, therefore, that temperature
would lead to a breaking of large gauge invariance in such a
system. This is, on the other hand, completely counter intuitive
considering that temperature should have no direct influence on gauge
invariance of the theory.

\subsection*{C-S Theory in $0+1$ Dimension}

As we have noted, Chern-Simons terms can exist in odd space-time
dimensions. Consequently, let us try to understand this puzzle of
large gauge invariance in a simple quantum mechanical theory. Let us
consider a simple theory of an interacting massive fermion with a
Chern-Simons term in $0+1$ dimension described by [14, 32]
\begin{equation}
L = \overline{\psi}_{j}(i\partial_{t} - A - m)\psi_{j} - \kappa
A\label{g6}
\end{equation}
Here, $j=1,2,\cdots,N_{f}$ labels the fermion flavors. There are
several things to note from this. First, we are considering an Abelian
gauge field for simplicity. Second, in this simple model, the gauge
field has no dynamics (in $0+1$ dimension the field strength is zero)
and, therefore, we do not have to get into the intricacies of gauge
theories. There is no Dirac matrix in $0+1$ dimension as well making
the fermion part of the theory quite simple as well. And, finally, the
Chern-Simons term, in this case, is a linear field so that we can, in
fact, think of the gauge field as an auxiliary field.

In spite of the simplicity of this theory, it displays a rich
structure including all the properties of the $2+1$ dimensional
theory that we have discussed earlier. For example, let us note that
under a gauge transformation
\begin{equation}
\psi_{j}\rightarrow e^{-i\lambda(t)}\psi_{j},\qquad A\rightarrow A +
\partial_{t}\lambda(t)\label{g7}
\end{equation}
the fermion part of the Lagrangian is invariant, but the Chern-Simons
term changes by a total derivative giving
\begin{equation}
S = \int dt\,L \rightarrow S -2\pi\kappa N\label{g8}
\end{equation}
where 
\begin{equation}
N = {1\over 2\pi} \int dt\,\partial_{t}\lambda(t)\label{g9}
\end{equation}
is the winding number and is an integer which vanishes for small gauge
transformations. Let us note that a large gauge transformation can have
a parametric form of the form, say,
\begin{equation}
\lambda(t) = -iN\,\log\left({1+it\over 1-it}\right)\label{g10}
\end{equation}
The fact that $N$ has to be an integer can be easily seen to arise
from the requirement of single-valuedness for the fermion field. Once
again, in light of our earlier discussion, it is clear from
eq. (\ref{g8}) that the theory is meaningful only if $\kappa$, the
coefficient of the Chern-Simons term, is an integer.

Let us assume, for simplicity, that $m>0$ and compute the correction
to the photon one-point function arising from the fermion loop at zero
temperature.
\begin{equation}
iI_{1} = -(-i)N_{f}\,\int {dk\over 2\pi}\,{i(k+m)\over
k^{2}-m^{2}+i\epsilon} = {iN_{f}\over 2}\label{g11}
\end{equation}
This shows that, as a result of the quantum correction, the
coefficient of the Chern-Simons term would change as
\[
\kappa\rightarrow \kappa - {N_{f}\over 2}
\]
As in $2+1$ dimensions, it is clear that the coefficient of the
Chern-Simons term would continue to be quantized and large gauge
invariance would hold if the number of fermion flavors is even. At
zero temperature, we can also calculate the higher point functions due
to the fermions in the theory and they all vanish. This has a simple
explanation following from the small gauge invariance of the
theory. Namely, suppose we had a nonzero two point function, then, it
would imply a quadratic term in the effective action of the form
\begin{equation}
\Gamma_{2} = {1\over 2} \int
dt_{1}\,dt_{2}\,A(t_{1})F(t_{1}-t_{2})A(t_{2})\label{g12}
\end{equation}
Furthermore, invariance under a small gauge transformation would imply
\begin{equation}
\delta\Gamma_{2} = -\int
dt_{1}\,dt_{2}\,\lambda(t_{1})\partial_{t_{1}}F(t_{1}-t_{2})A(t_{2}) =
0\label{g13}
\end{equation}
The solution to this equation is that $F=0$ so that there cannot be a
quadratic term in the effective action which would be local and yet be
invariant under small gauge transformations. A similar analysis would
show that small gauge invariance does not allow any higher point
function to exist at zero temperature. 

Let us also note that eq. (\ref{g13}) has another solution, namely,
\[
F(t_{1}-t_{2}) = {\rm constant}
\]
In such a case, however, the quadratic action becomes non-extensive,
namely, it is the square of an action. We do not expect such terms to
arise at zero temperature and hence the constant has to vanish for
vanishing temperature. As we will see next, the constant does not have
to vanish at finite temperature and we can have non-vanishing higher
point functions implying a non-extensive structure of the effective
action. 

The fermion propagator at finite temperature (in the real time
formalism) has the form [10]
\begin{eqnarray}
S(p) & = & (p+m)\left({i\over p^{2}-m^{2}+i\epsilon} -
2\pi n_{F}(|p|)\delta(p^{2}-m^{2})\right)\nonumber\\
 & = & {i\over p-m+i\epsilon} - 2\pi n_{F}(m)\delta(p-m)\label{g14}
\end{eqnarray}
and the structure of the effective action can be studied in the
momentum space in a straightforward manner. However, in this simple
model,  it is much easier to analyze the amplitudes in the coordinate
space.  Let us note that the coordinate space structure of the fermion
propagator is quite simple, namely,
\begin{equation}
S(t) = \int {dp\over 2\pi}\,e^{-ipt}\,\left({i\over p-m+i\epsilon} -
2\pi n_{F}(m)\delta(p-m)\right) =
(\theta(t)-n_{F}(m))e^{-imt}\label{g15}
\end{equation}
In fact, the calculation of the one point function is trivial now
\begin{equation}
iI_{1} = -(-i)N_{f}S(0) = {iN_{f}\over 2}\,\tanh {\beta m\over
2}\label{g16}
\end{equation}
This shows that the behavior of this theory is completely parallel to
the $2+1$ dimensional theory in that, it would suggest
\[
\kappa\rightarrow \kappa -{N_{f}\over 2}\,\tanh {\beta m\over 2}
\]
and it would appear that large gauge invariance would not hold at
finite temperature.

Let us next calculate the two point function at finite temperature.
\begin{eqnarray}
iI_{2} & = & -(-i)^{2}\,{N_{f}\over
2!}\,S(t_{1}-t_{2})S(t_{2}-t_{1})\nonumber\\
 & = & -{N_{f}\over 2}\,n_{F}(m)(1-n_{F}(m))\nonumber\\
 & = & -{N_{f}\over 8}\, {\rm sech}^{2} {\beta m\over 2} = {1\over
2}\,{1\over 2!}\,{i\over\beta}\,{\partial(iI_{1})\over \partial m}\label{g17}
\end{eqnarray}
This shows that the two point function is a constant as we had noted
earlier  implying that the quadratic term in the effective action
would be non-extensive.

Similarly, we can also calculate the three point function trivially
and it has the form
\begin{equation}
iI_{3} = {iN_{f}\over 24}\,\tanh {\beta m\over 2}\,{\rm sech}^{2} {\beta
m\over 2} = {1\over 2}\,{1\over
3!}\,\left({i\over\beta}\right)^{2}\,{\partial^{2}(iI_{1})\over
\partial m^{2}}\label{g18}
\end{equation}
In fact, all the higher point functions can be worked out in a
systematic manner. But, let us observe a simple method of computation
for these. We note that because of the gauge invariance (Ward
identity), the amplitudes cannot depend on the external time
coordinates as is clear from the  calculations of the lower point
functions. Therefore, we can always simplify the calculation by
choosing a particular time ordering convenient to us. Second, since we
are evaluating a loop diagram (a fermion loop) the initial and the
final time coordinates are the same and, consequently, the phase
factors in the propagator (\ref{g15}) drop out. Therefore, let us
define a simplified propagator without the phase factor as
\begin{equation}
\widetilde{S}(t) = \theta(t) - n_{F}(m)\label{g19}
\end{equation}
so that we have
\begin{equation}
\widetilde{S}(t>0) = 1 - n_{F}(m),\qquad S(t<0) = -
n_{F}(m)\label{g20}
\end{equation}
Then, it is clear that with the choice of the time ordering,
$t_{1}>t_{2}$, we can write
\begin{eqnarray}
{\partial\widetilde{S}(t_{1}-t_{2})\over \partial m} & = & -\beta
\widetilde{S}(t_{1}-t_{3})\widetilde{S}(t_{3}-t_{2})\qquad
t_{1}>t_{2}>t_{3}\nonumber\\
{\partial\widetilde{S}(t_{2}-t_{1})\over \partial m} & = & -\beta
\widetilde{S}(t_{2}-t_{3})\widetilde{S}(t_{3}-t_{1})\qquad
t_{1}>t_{2}>t_{3}\label{g21}
\end{eqnarray}

In other words, this shows that differentiation of a fermionic
propagator with respect to the mass of the fermion is equivalent to
introducing an external photon vertex (and, therefore, another fermion
propagator as well) up to constants. This is the analogue of the Ward identity
in QED in four dimensions except that it is much simpler. From this
relation, it is clear that if we take a $n$-point function and
differentiate this with respect to the fermion mass, then, that is
equivalent to adding another external photon vertex in all possible
positions. Namely, it should give us the $(n+1)$-point function up to
constants. Working out the details, we have,
\begin{equation}
{\partial I_{n}\over \partial m} = -i\beta(n+1)I_{n+1}\label{g22}
\end{equation}
Therefore, the $(n+1)$-point function is related to the $n$-point
function recursively and, consequently, all the amplitudes are
related to the one point function which we have already
calculated. (Incidentally, this is already reflected in
eqs. (\ref{g17},\ref{g18})). 

With this, we can now determine the full effective action of the
theory at finite temperature to be
\begin{eqnarray}
\Gamma & = & -i\,\sum_{n}\,a^{n}\,(iI_{n})\nonumber\\
 & = & -{i\beta N_{f}\over 2}\,\sum_{n}\,{(ia/\beta)^{n}\over
 n!}\,\left({\partial\over \partial m}\right)^{n-1}\,\tanh {\beta
 m\over 2}\nonumber\\
 & = & -iN_{f}\,\log \left(\cos {a\over 2} + i \tanh {\beta m\over
 2}\,\sin {a\over 2}\right)\label{g23}
\end{eqnarray}
where we have defined
\begin{equation}
a = \int dt\,A(t)\label{g24}
\end{equation}

There are several things to note from this result. First of all, the
higher point functions are no longer vanishing at finite temperature
and give rise to a non-extensive structure of the effective
action. More importantly, when we include all the higher point
functions, the complete effective action is invariant under large
gauge transformations, namely, under
\begin{equation}
a\rightarrow a + 2\pi N\label{g25}
\end{equation}
the effective action changes as 
\begin{equation}
\Gamma\rightarrow \Gamma + N N_{f}\pi\label{g26}
\end{equation}
which leaves the path integral invariant for an even number of fermion
flavors. This clarifies the puzzle of large gauge invariance at finite
temperature in this model. Namely, when we are talking about
large changes (large gauge transformations), we cannot ignore higher
order terms if they exist. This may provide a resolution to the large
gauge invariance puzzle in the $2+1$ dimensional theory as well. However, in
spite of several nice analysis [33], this puzzle has not yet been settled
in all its generality in the $2+1$ dimensional case.

\subsection*{Exact Result}

In the earlier section, we discussed a perturbative method of
calculating the effective action at finite temperature which clarified
the puzzle of large gauge invariance. However, this quantum mechanical
model is simple enough that we can also evaluate the effective action
directly and, therefore, it is worth asking how  the perturbative
calculations compare with the exact result.

The exact evaluation of the effective action can be done easily using
the imaginary time formalism. But, first, let us note that the
fermionic part of the Lagrangian in eq. (\ref{g6}) has the form
\begin{equation}
L_{f} = \overline{\psi}(i\partial_{t} - A - m)\psi\label{g27}
\end{equation}
where we have suppressed the fermion flavor index for simplicity. Let
us note that if we make a field redefinition of the form
\begin{equation}
\psi(t) = e^{-i\int_{0}^{t} dt'\,A(t')}\,\tilde{\psi}(t)\label{g28}
\end{equation}
then, the fermionic part of the Lagrangian becomes free, namely,
\begin{equation}
L_{f} = \overline{\tilde{\psi}}\,(i\partial_{t} -
m)\tilde{\psi}\label{g29}
\end{equation}

This is a free theory and, therefore, the path integral can be easily
evaluated. However, we have to remember that the field redefinition in
(\ref{g28})  changes the periodicity condition for the fermion
fields. Since the original fermion field was expected to satisfy 
anti-periodicity 
\[
\psi(\beta) = - \psi(0)
\]
it follows now that the new fields must satisfy
\begin{equation}
\tilde{\psi}(\beta) = - e^{-ia}\,\tilde{\psi}(0)\label{g30}
\end{equation}
Consequently, the path integral for the free theory (\ref{g29}) has to
be evaluated subject to the periodicity condition of (\ref{g30}).

Although the periodicity condition (\ref{g30}) appears to be
complicated, it is well known that the effect can be absorbed by
introducing a chemical potential [10], in the present case, of the form
\begin{equation}
\mu = {ia\over \beta}\label{g31}
\end{equation}
With the addition of this chemical potential, the path integral can be
evaluated  subject to the usual anti-periodicity condition. The
effective action can now be easily determined
\begin{eqnarray}
\Gamma & = & -i\,\log \left({\det
(i\partial_{t}-m+{ia\over\beta})\over
(i\partial_{t}-m)}\right)^{N_{f}}\nonumber\\
 & = & -iN_{f}\,\log \left({\cosh {\beta\over 2}(m-{ia\over \beta})\over
\cosh {\beta m\over 2}}\right)\nonumber\\
 & = & -iN_{f}\,\log \left(\cos {a\over 2} + i\tanh {\beta m\over
2}\,\sin {a\over 2}\right)\label{g32}
\end{eqnarray}
which coincides with the perturbative result of eq. (\ref{g24}).

\section{Supersymmetry Breaking}

One of the reasons for studying finite temperature field theory is to
understand questions such as phase transitions in such systems. It is
by now well understood that most field theoretic models of spontaneous
symmetry breaking display a phase structure much like what one sees in
a magnet, namely, above a certain critical temperature, the system is
in a symmetric phase while below the critical temperature, the system
is in a broken symmetry phase. Thus, temperature has the almost
universal effect that if a symmetry is spontaneously broken at low
temperature, it is restored  at  temperatures above a certain
critical value. Qualitatively, it can be understood as
follows. Temperature, particularly  high temperature, provides a
lot of thermal energy to a physical system to wash out any structure
in the zero temperature potential which may be responsible for
symmetry breaking. There is, however, one class of symmetries where
temperature has the inverse effect, namely, in a supersymmetric
theory, a symmetric phase at low temperature goes to a broken phase at
high temperature. (Of course, if supersymmetry is broken at low
temperature, it continues to be broken even at high temperature.) We
will discuss this phenomenon with a simple quantum mechanical model in
this section.

\subsection*{Supersymmetric Oscillator at $T=0$}

Let us note that supersymmetry is an ultimate form of symmetry that
one can dream of, namely, it transforms bosons into fermions and {\it
vice versa} [34-35]. To introduce supersymmetry, let us consider a simple
quantum mechanical model, commonly known as the supersymmetric
oscillator [16]. It consists of a bosonic and a fermionic oscillator of the
same frequency. Therefore, we can write the Hamiltonian, for the
system as
\begin{equation}
H = H_{B} + H_{F} = \omega \left(a_{B}^{\dagger}a_{B} +
a_{F}^{\dagger}a_{F}\right)\label{h1}
\end{equation}
where $a_{B}$ and $a_{F}$ describe, respectively, the bosonic and the
fermionic annihilation operators.

The immediate thing to note from the structure of the Hamiltonian in
eq. (\ref{h1}) is that there is no zero point energy. We will see this
shortly as a general feature of supersymmetric theories. Let us also
define two fermionic operators of the form
\begin{equation}
Q = a_{B}^{\dagger}a_{F},\qquad \overline{Q} =
a_{F}^{\dagger}a_{B}\label{h2}
\end{equation}
With the usual canonical commutation relations for the bosonic
operators (see eq. (\ref{d29})) and anti-commutation relations for the
fermionic operators (see eq. (\ref{d11})), it is easy to check that
\[
\left[Q , H\right] = 0 = \left[\overline{Q} , H\right]
\]
Namely, these fermionic operators are conserved. In fact, together
with the Hamiltonian, they satisfy the algebra (it is straightforward
to check this)
\begin{eqnarray}
\left[Q , H\right] & = & 0 = \left[\overline{Q} , H\right]\nonumber\\
\left[Q , Q\right]_{+} & = & 0 = \left[\overline{Q} ,
\overline{Q}\right]_{+}\nonumber\\ 
\left[Q , \overline{Q}\right]_{+} & = & {1\over \omega}\,H\label{h3}
\end{eqnarray}
Such an algebra, where both commutators and anti-commutators are
involved (or alternately, where there is a grading of the
multiplication rule of the algebra), is known as a graded Lie algebra
and supersymmetric theories are realizations of graded  Lie
algebras.

As we know from the study of symmetries, conserved quantities generate
infinitesimal symmetries of the theory. Since both $Q$ and
$\overline{Q}$ are conserved, it is worth asking what kind of symmetry
transformations of the theory they generate. In fact, let us keep in
mind that they are fermionic operators and hence the symmetry they
will generate cannot be conventional. Explicitly, we can check that
\begin{eqnarray}
\left[Q , a_{B}^{\dagger}\right] & = & 0 = \left[Q ,
a_{F}\right]_{+}\nonumber\\ 
\left[Q , a_{B}\right] & = & -a_{F}\nonumber\\
\left[Q , a_{F}^{\dagger}\right]_{+} & = & a_{B}^{\dagger}\nonumber\\
\left[\overline{Q} , a_{B}\right] & = & 0 = \left[\overline{Q} ,
a_{F}^{\dagger}\right]_{+}\nonumber\\
\left[\overline{Q} , a_{B}^{\dagger}\right] & = &
a_{F}^{\dagger}\nonumber\\
\left[\overline{Q} , a_{F}\right]_{+} & = & a_{B}\label{h4}
\end{eqnarray}
Namely, $Q$ and $\overline{Q}$ take bosonic operators to fermionic
ones and {\it vice versa} which is the bench mark of
supersymmetry. Thus, our Hamiltonian in eq. (\ref{h1}) is invariant
under supersymmetric transformations of the form (\ref{h4}).

There are several things to note from the structure of the
supersymmetry algebra in eq. (\ref{h3}). First, the energy
eigenvalues of our supersymmetric theory have to be positive
semi-definite since the operator on the left hand side of the last
relation in (\ref{h3}) is. Furthermore, if the ground state is
supersymmetric satisfying
\begin{equation}
Q|0\rangle = 0 = \overline{Q}|0\rangle\label{h5}
\end{equation}
then, the ground state will have vanishing energy, as we had pointed
out earlier as the case for our system. Both these results are, in
fact, quite general for any supersymmetric theory. We also note from
the structure of the algebra that the spectrum of the Hamiltonian will
be doubly degenerate except for the ground state. Namely, if
$|\psi\rangle$ is an eigenstate of the Hamiltonian, then,
$Q|\psi\rangle$ (or, $\overline{Q}|\psi\rangle$ -- only one of them
would be nontrivial depending on the form of $|\psi\rangle$) would
be degenerate in energy.

Let us, in fact, examine some of these general results explicitly. The
spectrum of the Hamiltonian in eq. (\ref{h1}) is, in fact, quite
straightforward. The Hilbert space is a product space containing
bosonic and fermionic oscillator states and a general state
has the structure
\begin{equation}
|n_{B},n_{F}\rangle = |n_{B}\rangle\otimes |n_{F}\rangle =
 {(a_{B}^{\dagger})^{n_{B}} (a_{F}^{\dagger})^{n_{F}}\over
 \sqrt{n_{B}!}}|0,0\rangle\label{h6}
\end{equation}
with energy eigenvalues
\begin{equation}
E_{n_{B},n_{F}} = \omega (n_{B}+n_{F}),\qquad
n_{F}=0,1;\;n_{B}=0,1,2,\cdots\label{h7}
\end{equation}
where the ground state is expected to satisfy
\begin{equation}
a_{B}|0\rangle = 0 = a_{F}|0\rangle\label{h8}
\end{equation}
We note that an immediate consequence of (\ref{h8}) is that
\[
Q|0\rangle = 0 = \overline{Q}|0\rangle
\]
and, consequently, the ground state is supersymmetric and that the
ground state energy is seen from (\ref{h7}) to vanish. All the higher
states have positive energy. Furthermore, we note that all the states
(except the ground state) of the form $|n_{B},0\rangle$ and
$|n_{B}-1,1\rangle$ are degenerate in energy. Let us also note the
effect of $Q$ and $\overline{Q}$ acting on the states of the Hilbert
space, namely,
\begin{eqnarray}
Q|n_{B},n_{F}\rangle & = & \left\{\begin{array}{cl}
\sqrt{n_{B}+1}\,|n_{B}+1,n_{F}-1\rangle & {\rm if}\;n_{F}\neq 0\\
0 & {\rm if}\;n_{F}=0
\end{array}\right.\nonumber\\
\overline{Q}|n_{B},n_{F}\rangle & = & \left\{\begin{array}{cl}
{1\over\sqrt{n_{B}}}\,|n_{B}-1,n_{F}+1\rangle & {\rm if}\;n_{B}\neq 0\,
{\rm or},\,n_{F}\neq 1\\
0 & {\rm if}\;n_{B}=0\,{\rm or},\,n_{F}=1
\end{array}\right.\label{h9}
\end{eqnarray}

\subsection*{Supersymmetric Oscillator at $T\neq 0$}

Let us next analyze the supersymmetric oscillator at finite
temperature in the formalism of thermo field dynamics. As we had noted
earlier, this is the ideal setting to discuss questions such as
symmetry breaking. Let us note, even before carrying out the
calculations, that we expect supersymmetry to be broken at finite
temperature. Intuitively, this is quite clear. Namely, supersymmetry
takes bosons to fermions and {\it vice versa} and, consequently, any
boundary condition that distinguishes between the two would lead to a
breaking of this symmetry. Temperature, in fact, introduces such a
condition, namely, bosons and fermions behave differently at finite
temperature (they obey distinctly different statistics). However, what
is not clear {\it a priori} is whether such a breaking would be
explicit or spontaneous.

To study the system at finite temperature within the framework of
thermo field dynamics, let us look at the complete system, including the
tilde oscillators, described by 
\begin{equation}
\hat{H} = H - \widetilde{H} =
\omega(a_{B}^{\dagger}a_{B}+a_{F}^{\dagger}a_{F}) -
\omega(\tilde{a}_{B}^{\dagger}\tilde{a}_{B} +
\tilde{a}_{F}^{\dagger}\tilde{a}_{F})\label{h10}
\end{equation}
The Hilbert space of the doubled system has the structure
\begin{equation}
|n_{B},n_{F}; \tilde{n}_{B},\tilde{n}_{F}\rangle =
 |n_{B},n_{F}\rangle\otimes
|\tilde{n}_{B},\tilde{n}_{F}\rangle\label{h11}
\end{equation}

The thermal vacuum can now be defined (as discussed in section {\bf
3}). Let us define
\[
G(\theta_{B},\theta_{F}) =
-i\theta_{B}(\beta)(\tilde{a}_{B}a_{B}-a_{B}^{\dagger}\tilde{a}_{B}^{\dagger})
- i\theta_{F}(\beta)(\tilde{a}_{F}a_{F} -
a_{F}^{\dagger}\tilde{a}_{F}^{\dagger})
\]
with (see eqs. (\ref{d22})and (\ref{d35}))
\begin{equation}
\tan\theta_{F}(\beta) = e^{-\beta\omega/ 2} =
\tanh\theta_{B}(\beta)\label{h12}
\end{equation}
Then, the thermal vacuum can be defined as
\begin{equation}
|0,\beta\rangle = e^{-iG(\theta_{B},\theta_{F})}\,|0\rangle\label{h13}
\end{equation}
This also allows us to calculate the thermal operators in a
straightforward manner. 

Let us note next that the expectation value of the Hamiltonian in the
thermal vacuum is given by
\begin{eqnarray}
E_{0}(\beta) & = & \langle 0,\beta|H|0,\beta\rangle = \langle
0,\beta|\omega(a_{B}^{\dagger}a_{B}+a_{F}^{\dagger}a_{F})|0,\beta\rangle\nonumber\\
 & = & \omega (\sinh^{2}\theta_{B}(\beta)+\sin^{2}\theta_{F}(\beta)) =
{2\omega e^{-\beta\omega}\over (1-e^{-2\beta\omega})}\label{h14}
\end{eqnarray}
This shows that the energy of the thermal vacuum is nonzero for any
finite temperature signaling that supersymmetry is
broken. Furthermore, let us note that
\begin{eqnarray}
Q|0,\beta\rangle & = & a_{B}^{\dagger}a_{F}|0,\beta\rangle =
{e^{-\beta\omega/2}\over\sqrt{1-e^{-2\beta\omega}}}|n_{B}(\beta)=1,n_{F}(\beta)=0;\tilde{n}_{B}(\beta)=0,\tilde{n}_{F}(\beta)=1\rangle\nonumber\\
\overline{Q}|0,\beta\rangle & = & a_{F}^{\dagger}a_{B}|0,\beta\rangle
= {e^{-\beta\omega/2}\over\sqrt{1-e^{-2\beta\omega}}}|n_{B}(\beta)=0,n_{F}
(\beta)=1;\tilde{n}_{B}(\beta)=1,\tilde{n}_{F}(\beta)=0\rangle\label{h15}
\end{eqnarray}
This, in fact, shows that supersymmetry breaking is spontaneous at
finite temperature and the new states on the right hand side of
(\ref{h15}) would correspond to the appropriate quasi particle
Goldstino  states associated with such a symmetry breaking.

There are various other order parameters for the breaking of
supersymmetry and all of them lead to the same conclusion that
supersymmetry is spontaneously broken at finite temperature [16].

\section{Conclusion}

In this article, we have tried to describe some of the interesting
features of finite temperature field theories. There are, of course,
many more topics that we have not been able to discuss. However, it is
our hope that the topics discussed, in this article, would raise the
curiosity of the readers to pursue  various other questions in this field.

This work was supported in part by the U.S. Dept. of
Energy Grant  DE-FG 02-91ER40685.


\begin{thebibliography}{99}
\bibitem{l1} T. Matsubara, {\it Prog. Theor. Phys.} {\bf 14} (1955) 351.
\bibitem{l2} J. Schwinger, {\it J. Math. Phys.} {\bf 2} (1961) 407;
J. Schwinger, {\it Lecture Notes Of Brandeis University Summer
Institute} (1960).
\bibitem{l3} H. Umezawa, H.Matsumoto and M. Tachiki, {\it Thero Field
Dynamics and Condensed States}, North-Holland, Amsterdam,1982.
\bibitem{l4} D. A. Kirzhnits and A. D. Linde, {\it Phys. Lett.} {\bf
42B} (1979) 471; L. Dolan and R. Jackiw, {\it Phys. Rev.} {\bf D9}
(1974) 3320; S. Weinberg, {\it Phys. Rev.} {\bf D9} (1974) 3357. 
\bibitem{l5} D. J. Gross, R. D. Pisarski and L. G. Yaffe, {\it
Rev. Mod. Phys.} {\bf 53} (1981) 43.
\bibitem{l6} A. A. Anselm, {\it Phys. Lett.} {\bf B217} (1989) 169;
A. A. Anselm and M. G. Ryskin, {\it Phys. Lett.} {\bf B266} (1991)
482; J. D. Bjorken, {\it Int. J. Mod. Phys.} {\bf A7} (1992) 4189;
J. P. Blaizot and A. Krzywicki, {\it Phys. Rev.} {\bf D46} (1992) 246;
K. Rajagopal and F. Wilczek, {\it Nuc. Phys.} {\bf B204} (1993) 577;
P. F. Bedaque and A. Das, {\it Mod. Phys. Lett.} {\bf A8} (1993) 3151.
\bibitem{l7} A. Das, {\it Field Theory, A Path Integral Approach}, World
Scientific (1993).
\bibitem{l8} A. L. Fetter and J. D. Walecka, {\it Quantum Theory of Many
Particle Systems}, McGraw-Hill (1971); A. A. Abrikosov, L. P. Gorkov
and I. E. Dzyaloshinski, {\it Methods of Quantum Field Theory in
Statistical Physics}, Dover (1975).
\bibitem{l9} J. I. Kapusta, {\it Finite Temperature Field Theory},
Cambridge University Press (1989); M. Le Bellac, {\it Thermal Field
Theory}, Cambridge University Press (1996).
\bibitem{l10} A. Das, {\it Finite Temperature Field Theory}, World
Scientific (1997).
\bibitem{l11} N. P. Landsman and C. G. van Weert, {\it Phys. Rep.} {\bf
145} (1987) 141.
\bibitem{l12} P. M. Bakshi and K. T. Mahanthappa, {\it J. Math. Phys.}
{\bf 4} (1963) 1; L. V. Keldysh, {\it Sov. Phys. JETP} {\bf 20} (1965)
1018; K. C. Chou et al, {\it Phys. Rep.} {\bf 118} (1985) 1.
\bibitem{l13} H. A. Weldon, {\it Phys. Rev.} {\bf D47} (1993) 594;
P. F. Bedaque and A. Das, {\it Phys. Rev.} {\bf D47} (1993) 601.
\bibitem{l14} G. Dunne, K. Lee and C. Lu, {\it Phys. Rev. Lett.} {\bf 78}
(1997) 3434.
\bibitem{l15} A. Das and G. Dunne, {\it Phys. Rev.} {\bf D57} (1998) 5023.
\bibitem{l16} A. Das and M. Kaku, {\it Phys. Rev.} {\bf D18} (1978) 4540;
A. Das, A. Kharev and V. S. Mathur, {\it Phys. Lett.} {\bf B181}
(1986) 299; A. Das and V. S. Mathur, {\it Phys. Rev.} {\bf D35} (1987)
2053; A. Das, {\it Physica} {\bf A158} (1989) 1.
\bibitem{l17} R. P. Feynman and A. R. Hibbs, {\it Quantum Mechanics and
Path Integrals}, McGraw-Hill (1965). 
\bibitem{l18} L. S. Shulman, {\it Techniques and Applications of Path
Integration}, John-Wiley (1981); H. Kleinert, {\it Path Integrals},
World Scientific (1995).
\bibitem{l19} F. A. Berezin, {\it The Method of Second Quantization},
Academic Press (1966); B. DeWitt, {\it Supermanifolds}, Cambridge
University Press (1984).
\bibitem{l20} R. Kubo, {\it J. Phys. Soc. Japan}, {\bf 12} (1957) 570;
P. Martin and J. Schwinger, {\it Phys. Rev.} {\bf 115} (1959) 1342.
\bibitem{l21} C. Bloch, {\it Nuc. Phys.} {\bf 7} (1958) 451.
\bibitem{l22} A. A. Abrikosov, L. P. Gorkov and I. E. Dzyaloshinski, {\it
Sov. Phys. JETP} {\bf 9} (1959) 636; H. Umezawa, Y. Tomozawa and
H. Ezawa, {\it Nuovo Cim.} {\bf 5} (1957) 810.
\bibitem{l23} H. A. Weldon, {\it Phys. Rev.} {\bf D26} (1982) 1394; {\it
ibid} {\bf D28} (1983) 2007.
\bibitem{l24} G. Baym and N. Mermin, {\it J. Math. Phys.} {\bf 2} (1961)
232.
\bibitem{l25} Y. Takahashi and H. Umezawa, {\it Collective Phenomena}
{\bf 2} (1975) 55.
\bibitem{l26} I. Ojima, {\it Ann. Phys.} {\bf 137} (1981) 1.
\bibitem{l27} A. J. Niemi and G. Semenoff, {\it Ann. Phys.} {\bf 152}
(1984) 105.
\bibitem{l28} P. F. Bedaque and A. Das, {\it Phys. Rev.} {\bf D45} (1992)
2906.
\bibitem{l29} A. Das and M. Hott, {\it Mod. Phys. Lett.} {\bf A9} (1994) 3383.
\bibitem{l30} S. Deser, R. Jackiw and S. Templeton, {\it Ann. Phys.} {\bf
140} (1982) 372.
\bibitem{l31} K. S. Babu, A. Das and P. Panigrahi, {\it Phys. Rev.} {\bf
D36} (1987) 3725; I. Aitchison and J. Zuk, {\it Ann. Phys.} {\bf 242}
(1995) 77; N. Brali\'{c}, C. Fosco and F. Schaposnik, {\it
Phys. Lett.} {\bf B383} (1996) 199; D. cabra, E. Fradkin, G. Rossini
and F. Schaposnik, {\it Phys. Lett.} {\bf B383} (1996) 434.
\bibitem{l32} G. Dunne, R. Jackiw and C. Trugenberger, {\it Phys. Rev.}
{\bf D41} (1990) 661.
\bibitem{l33} S. Deser, L. Griguolo and D. Seminara, {\it
Phys. Rev. Lett.} {\bf 79} (1997) 1976; C. Fosco, G. Rossini and
F. Schaposnik, {\it Phys. Rev.} {\bf 79} (1997) 1980; S. Deser,
L. Griguolo and D. Seminara, {\it Phys. Rev.} {\bf D57} (1998) 7444;
C. Fosco, G. Rossini and F. Schaposnik, {\it Phys. Rev.} {\bf D56}
(1997) 6547.
\bibitem{l34} Y. A. Gel'fand and E. P. Likhtman, {\it JETP Lett.} {\bf
13} (1971) 323; P. Ramond, {\it Phys. Rev.} {\bf D3} (1971) 2415;
A. Neveu and J. Schwarz, {\it Nuc. Phys.} {\bf 31} (1971) 86;
D. Volkov and V. Akulov, {\it Phys. Lett.} {\bf B46} (1974) 109;
J. Wess and B. Zumino, {\it Nuc. Phys.} {\bf 70} (1974) 39.
\bibitem{l35} P. Fayet and S. Ferrara, {\it Phys. Rep.} {\bf 32} (1977)
249; M. F. Sohnius, {\it Phys. Rep.} {\bf 128} (1985) 39.
\end{thebibliography}
\end{document}